\documentclass[preprint,showpacs,preprintnumbers,amsmath,amssymb]{revtex4}
\usepackage{graphicx}
\usepackage{dcolumn}
\usepackage{bm}
\usepackage{marvosym,color}

\begin{document}

\title{Resonant Auger decay of the core-excited C$^\ast$O molecule in intense X-ray laser fields}

\author{\firstname{Philipp~V.} \surname{Demekhin}}
\email{philipp.demekhin@pci.uni-heidelberg.de}
\altaffiliation[\\On leave from: ]{Rostov State Transport University,
Narodnogo Opolche\-niya square 2, 344038, Rostov-on-Don, Russia}
\affiliation{Theoretische Chemie, Physikalisch-Chemisches
Institut, Universit\"{a}t  Heidelberg, Im Neuenheimer Feld 229,
D-69120 Heidelberg, Germany}

\author{\firstname{Ying--Chih} \surname{Chiang}}
\affiliation{Theoretische Chemie, Physikalisch-Chemisches Institut, Universit\"{a}t Heidelberg,
Im Neuenheimer Feld 229, D-69120 Heidelberg, Germany}

\author{\firstname{Lorenz~S.} \surname{Cederbaum}}
\affiliation{Theoretische Chemie, Physikalisch-Chemisches Institut, Universit\"{a}t Heidelberg,
Im Neuenheimer Feld 229, D-69120 Heidelberg, Germany}

\date{\today}

\begin{abstract}
The dynamics of the resonant Auger (RA) process of the core-excited C$^\ast$O(1s$^{-1}\pi^\ast,v_r=0$) molecule in an intense X-ray laser field is studied theoretically. The theoretical approach includes the analogue of the conical intersections of the complex potential energy surfaces of the ground and `dressed' resonant states due to intense X-ray pulses, taking into account the decay of the resonance and the direct photoionization of the ground state, both populating the same final ionic states coherently, as well as the direct photoionization of the resonance state itself. The light-induced non-adiabatic effect of the analogue of the conical intersections of the resulting complex potential energy surfaces gives rise to strong coupling between the electronic, vibrational and rotational degrees of freedom of the diatomic CO molecule. The interplay of the direct photoionization of the ground state and of the decay of the resonance increases dramatically with the field intensity. The coherent population of a final ionic state via both the direct photoionization and the resonant Auger decay channels induces strong interference effects with distinct patterns in the RA electron spectra. The individual impact of these physical processes on the total electron yield and on the CO$^+(A~^2\Pi)$ electron spectrum are demonstrated. 
\end{abstract}


\maketitle

\section{Introduction}
\label{sec:intro}

Recently, the resonant Auger (RA) decay of atoms exposed to strong X-ray laser pulses has been studied theoretically \cite{Rohringer08,Liu10,Sun10,Demekhin11SFatom}. It has been demonstrated that under the extreme field conditions provided by X-ray Free Electron Lasers (XFELs) \cite{XFEL1,XFEL2}, such as unprecedented high-intensity, $\sim 10^{18}~\mathrm{W/cm}^{2}$, and very short pulse durations, 10--500~fs, the stimulated emission from the resonance back to the ground state starts to compete with the Auger decay. The interplay between the resonant excitation and stimulated emission results in Rabi flopping between the ground state and the resonance within its Auger decay lifetime and pulse duration. This leads to spectacular modifications of the RA spectra, like the appearance of multipeak structure in the electron spectra \cite{Rohringer08,Liu10}. It has also been demonstrated \cite{Liu10,Sun10,Demekhin11SFatom}, that direct photoionizations of the ground state and of the resonant state increase dramatically with the field intensity, resulting in additional leakages of the corresponding populations into all possible final ionic states. Moreover,  a final ionic state is populated coherently by both the direct photoionization from the ground state and by the resonant Auger decay inducing strong interference effects with distinct patterns in the RA electron spectra of atoms \cite{Demekhin11SFatom}.

What happens when a diatomic molecule is exposed to a strong laser pulse with a carrier frequency which fits to the energy difference between two electronic states? It has been demonstrated that the two resulting `dressed' electronic states exhibit a conical intersection in the space of the rotational angle and the internuclear distance \cite{Moiseyev08,Sindelka11}. The emergence of this light-induced conical intersection leads to substantial non-adiabatic effects and to a strong mixing of rotations and vibrations \cite{Sindelka11}. Very recently \cite{MolRaSfPRL}, the RA process in molecules in intense X-ray laser fields has been studied theoretically. Also here it has been demonstrated that an atomic-like picture (fixed internuclear distance and molecular orientation)  is inapplicable. Not only the vibrational motion plays a role in the decay as in the case of weak fields, but the rotational motion becomes very important in strong fields as well. In particular, the appearance of the analogue of conical intersections of the now complex (owing to a finite life-time of the electronic decaying state) potential energy surfaces (PESs) of the ground and `dressed' resonant states has been predicted in \cite{MolRaSfPRL}. This phenomenon gives rise to strong non-adiabatic coupling and to strong mixing between the electronic, vibrational and rotational degrees of freedom of diatomic molecules exposed to strong X-ray laser pulses. To demonstrate the non-adiabatic effects, the resonant ionization channel (i.e., excitation and decay of the resonance) has been explicitly included in the numerical calculations of the RA process in the HCl molecule studied in Ref.~\cite{MolRaSfPRL}, whereas the leakages and interference effects mentioned above for atoms were not taken into account explicitly for transparency of presentation.

In the present work we study how the combined action of all the aforementioned competing processes evoked by a strong field influence the multidimensional dynamics of the RA effect in diatomic molecules exposed to strong X-ray laser pulses. For this purpose we unify the previously developed theoretical approaches for RA in atoms   \cite{Demekhin11SFatom} and for molecules \cite{MolRaSfPRL}. As demonstrated in Ref.~\cite{Demekhin11SFatom}, the simultaneous effect of the RA decay and of the direct ionization of the involved states results in additional modifications of the Hamiltonian governing the nuclear dynamics on the coupled dressed surfaces. One can expect that these modifications will even further enhance the non-adiabatic effects caused by the appearance of the analogue of a conical intersection of the complex PESs. In our previous study \cite{MolRaSfPRL}, the case of excitation of the dissociative  HCl$^\ast(2p^{-1}\sigma^\ast)$ resonant state and its subsequent RA decay into one of the dissociative final states has been investigated. There, the nuclear dynamics accompanying the RA decay result in the fragmentation of of the HCl molecule, and, as a consequence, a rather simple RA spectrum appears. The RA decay of the HCl$^\ast(2p^{-1}\sigma^\ast)$   state is however difficult to study experimentally, owing to many overlapping intermediate and final electronic states.  In the present work we concentrate on the nuclear dynamics accompanying the RA decay of a bound resonant electronic state into a bound final ionic state of a diatomic molecule, where the corresponding vibrational structures can be resolved in the excitation and decay spectra.

The RA decay of the core-excited C$^\ast$O(1s$^{-1}\pi^\ast$) molecule is a perfect candidate for our purposes. First of all, for weak fields it is one of the most thoroughly studied processes both, experimentally \cite{Truesdale84,Hemmers93,BeckShirl,Pruemper08,Carravetta97,Piancastelli97,Osborne98,Kukk99,Demekhin09CstarO} and theoretically \cite{Carravetta97,Piancastelli97,Osborne98,Kukk99,Demekhin09CstarO,Gortel98,Bonhoff99,Fink09}. Therefore, all parameters for the calculations are available. Second, the weak field excitation spectrum of the core-excited resonance consists of three well resolved vibrational states $v_r =0,1$ and 2 with excitation probabilities of about 87\%, 12\% and 1\%, respectively \cite{Piancastelli97,Demekhin09CstarO}. The energy separation of  $\omega_e = 250$~meV \cite{Piancastelli97} between the $v_r$ levels of the C$^\ast$O resonance, is about three times larger than their natural lifetime width of $\Gamma_{aug}=80$~meV \cite{Prince99}, and thus the underlying so called lifetime vibrational interference (LVI, \cite{GelMukhanov77}) effects are small. The timescale of the nuclear vibrational motion, $\tau_v=2\pi/\omega_e\approx 16.5$~fs, and that of the RA decay, $\tau_d=1/\Gamma_{aug}\approx 8.2$~fs, are rather comparable. All of these imply that one can expect measurable fingerprints of the nuclear dynamics in the core-excited state when studying strong field effects. Finally, the RA electron spectra corresponding to the decay of the resonance to the three lowest $X,~A$, and $B$ states of the CO$^+$ ion do not overlap (the energy separations are about  2~eV \cite{BeckShirl,Kukk99}), and can thus be studied separately. In the present work we investigate the RA decay into the CO$^+(A~^2\Pi)$ final ionic state which exhibits the richest vibrational structure of all the electron spectra in the weak field case \cite{BeckShirl,Kukk99}.

\section{Theory}
\label{sec:thy}

\begin{figure}
\includegraphics[scale=0.55]{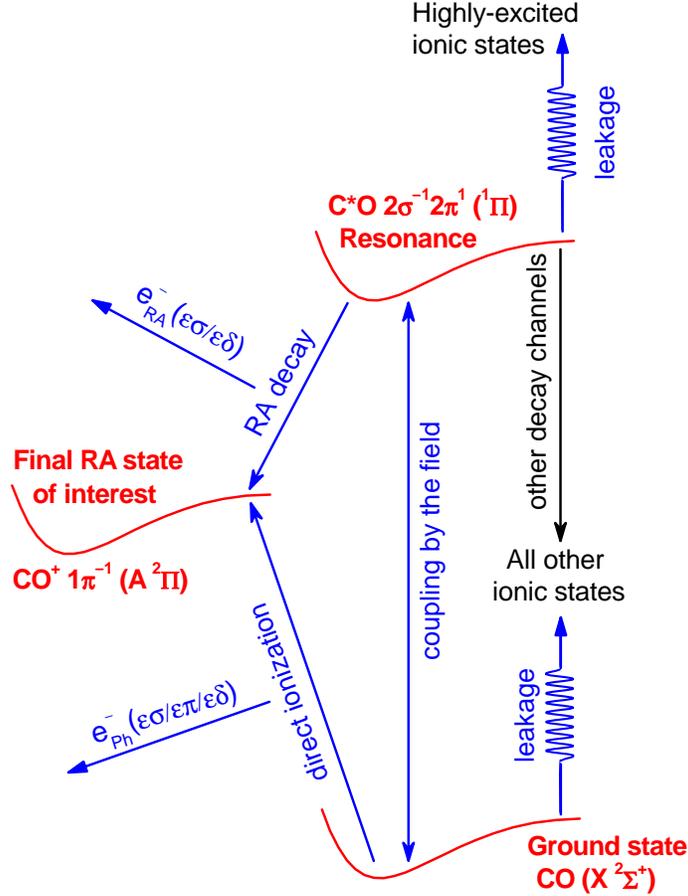}
\caption{(Color online) Schematic representation of   RA effect of the CO molecule exposed to strong X-ray pulses (see text for details).}
\label{fig:scheme}
\end{figure}

The processes relevant to the present study are represented schematically in Fig.~\ref{fig:scheme}. The strong X-ray laser field with the photon energy $\omega$ around 287.4~eV  couples the ground  and the resonance states. As demonstrated in Ref.~\cite{MolRaSfPRL}, this coupling leads to damped Rabi oscillations of the nuclear wave packets between the ground state and the resonance. The  CO$^+(A~^2\Pi)$  final ionic state can be coherently populated  via the participator Auger decay of the resonance and  by the direct photoionization from the ground state. The first channel is operative at those times when the wave packet is in the resonant state, and the second channel -- when it is in the ground state. In addition, there are two mechanisms of losses relevant to the RA effect in strong fields. The  total photoionizations of the ground state and of the resonance lead to  leakages of the respective wave packets, as indicated in Fig.~\ref{fig:scheme} by the vertical helix arrows from the ground state and from the resonance, respectively.  These processes, except of the Auger decay itself,  are operative only when the pulse is on.  After the pulse expires, the final population of the ground state can be much smaller than 1. In the presence of a strong pulse, multiple ionization of a molecule takes place as well  \cite{XFEL2}. Different possibilities for that were discussed in  Ref.~\cite{Demekhin11SFatom}.  However,  multiple ionization processes, in principle, can be experimentally separated from the RA effect as soon as one measures energies of ejected electrons and does not intend to measure ions.

The currently operating XFEL, Linac Coherent Light Source (LCLS), do not produce so far a monochromatic radiation and the  X-ray pulses consist of many spikes with random fluctuations of the frequency, phase and amplitude  \cite{XFEL1,XFEL2}. The impact of these problems on the RA effect of atoms has been studied in \cite{Rohringer08}. In addition, the X-ray pulse is reshaped during the propagation through a resonant medium \cite{Liu10,Sun10}. In order to reveal the individual contributions of all included physical processes, we concentrate here on the physics a single molecule undergoes when exposed to a coherent and monochromatic X-ray pulse.  For simplicity we assume a Gaussian-shaped pulse of   duration $\tau$ centered at $t_0$ ($g(t)=e^{-(t-t_0)^2/\tau^2}$) with linear polarization of the field along the $z$ axis:
\begin{equation}
\label{eq:e_vector}
\mathcal{E}(t)=\mathcal{E}_0(t)\cos\omega t=\mathcal{E}_0 \,g(t) \cos\omega t.
\end{equation}
Here $\mathcal{E}_0$ is the peak amplitude, and the pulse-shape function $g(t)$   varies slowly on the timescale of $2\pi/\omega$. The cycle-averaged intensity of the field is given in atomic units via (1 a.u. = 6.43641$\times 10^{15}$ W/cm$^2$)
\begin{equation}
\label{eq:intens}
I(t)=\frac{1}{8\pi\alpha}\left\{\mathcal{E}_0 \,g(t)\right\}^2 ,
\end{equation}
where $ \alpha=1/137.036$ is the fine structure constant.  For brevity of presentation, the following terminology is used throughout the manuscript. A pulse of a peak intensity well below 1 a.u. (6.43641$\times 10^{15}$ W/cm$^2$) is referred to hereafter as a weak pulse. A weak field does not induce noticeable Rabi oscillations (leads to negligibly small stimulated emission) and can be treated perturbatively. Pulses with peak intensities comparable to or larger than 1 a.u. are referred to hereafter as strong pulses. Strong fields govern photon absorption processes in the extremely nonlinear regime, e.g. by Rabi flopping \cite{Rohringer08,Demekhin11SFatom}.

To describe the dynamics of the RA effect theoretically as a function of time, one needs to solve the time-dependent Schr\"{o}dinger equation for the molecule and its interaction with the field  (atomic units $e=m_e=\hbar=1$ are used throughout)
\begin{equation}
\label{eq:hamilt}
i\dot{\Psi}(t)=\hat{H}(t)\Psi(t)=\left(\hat{H}_{nuc}+\hat{H}_{el}+ \hat{D}\, \mathcal{E}(t)\right)\Psi(t) . 
\end{equation} 
where $\hat{D}$ represents the dipole transition operator. The present approach to solve Eq.~(\ref{eq:hamilt})  combines those reported in Refs.~\cite{Demekhin11SFatom,MolRaSfPRL,Pahl99ZPD}.

\subsection{Nuclear dynamics Hamiltonian}
\label{sec:thyND}

We start with the randomly oriented rotating CO molecule in its ground ro-vibronic state CO$(X~^1\Sigma^+,v_0=0,J_0=0)$, i.e., with an isotropic  distribution over the rotational angle. We shall demonstrate below that the non-adiabatic effects of coupling the vibrational and rotational degrees of freedom of diatomic molecules induced by the laser field will change this distribution dramatically. Following Ref.~\cite{Pahl99ZPD} the total wave function of a rotating molecule as a function of time can be represented via the following ansatz including three electronic states: the ground state and the resonance with the electronic wave functions $\Phi_I$ and  $\Phi_R$, respectively, and the final ionic state CO$^+(A~^2\Pi)$ plus outgoing  electron of energy $\varepsilon$ with the total electronic wave function $\Phi^\varepsilon_{A}$:
\begin{equation}
\label{eq:anzatz}
\Psi(t)= \vert\Psi_I(t)\rangle \Phi_I +\vert\widetilde{\Psi}_R(t)\rangle \Phi_R +\int \vert \widetilde{\Psi}_A (\varepsilon,t) \rangle \Phi^\varepsilon_{A}  d\varepsilon.
\end{equation}
Here $\vert\Psi_I(t)\rangle$, $\vert\widetilde{\Psi}_R(t)\rangle$, and $\vert \widetilde{\Psi}_A (\varepsilon,t) \rangle$ are the time-dependent wave packets propagating on the PESs of the ground, resonant and final ionic states, respectively. It should be remembered, that these wave packets depend explicitly on the nuclear  coordinates $R$ and $\theta$. For brevity, we will explicitly  show the  dependence on the nuclear coordinates only in the final Hamiltonian matrix (\ref{eq:fin2}) governing the nuclear dynamics.

In order to obtain the final set of equations for the propagation of the nuclear wave packets, we substitute the ansatz  (\ref{eq:anzatz}) in the time-dependent  Schr\"{o}dinger equation (\ref{eq:hamilt}) and project the result onto each electronic state. Similarly to Refs.~\cite{Demekhin11SFatom,MolRaSfPRL,Pahl99ZPD}, we imply the rotating wave approximation \cite{Griffiths94}, the local approximation \cite{Cederbaum81,Domcke91}, and redefine (`dress') the time dependent wave packets of the resonant and final ionic states as follows
\begin{equation}
\label{eq:redefine}
\vert {\Psi}_R(t)\rangle= \vert\widetilde{\Psi}_R(t)\rangle\,e^{+i\omega t}~~~ \mathrm{and}~~~  \vert {\Psi}_A (\varepsilon,t) \rangle= \vert \widetilde{\Psi}_A (\varepsilon,t) \rangle\,e^{+i\omega t}.
\end{equation}
Utilizing the rotating wave and the local approximations is well justified for high energy resonant processes considered here \cite{Rohringer08,Demekhin11SFatom}. Details of the derivation can be found in Refs.~\cite{Demekhin11SFatom,MolRaSfPRL,Pahl99ZPD}.

In order to derive the final Hamiltonian matrix for the nuclear dynamics accompanying the studied process, we notice that the final ionic state  CO$^+(A~^2\Pi)$ can be populated via Auger decay of the resonance by the emission of either $\varepsilon \sigma$ or $\varepsilon \delta$ RA electrons (as indicated  in Fig.~\ref{fig:scheme}).  These channels form the  $^1\Pi$ symmetry  of the overall final state of the ion plus RA electron which is   the symmetry of the resonant state. Contrary to that, the direct population of the  CO$^+(A~^2\Pi)$ state from the ground state  may proceed via the emission of $\varepsilon \sigma$,  $\varepsilon \pi$ or $\varepsilon \delta$ photoelectrons (see  Fig.~\ref{fig:scheme}). The direct photoionization channels with $\varepsilon \sigma$  and $\varepsilon \delta$ photoelectrons contribute to the same overall final state as the resonant channels, which thus superimpose and interfere. Emission of $\varepsilon \pi$ photoelectrons, however, results in the  overall final state of  $^1\Sigma^+$ symmetry, which cannot be populated resonantly. Moreover, as will be shown below, these transitions are operative via different components of the  dipole transition operator  $\hat{D}$. That is why  it is important to separate the  final states explicitly  in the ansatz (\ref{eq:anzatz}) (the last integral in Eq.~(\ref{eq:anzatz}) includes both types of   final states).

Let us collect the individual  nuclear wave packets contributing to the total wave function (\ref{eq:anzatz}) into a single vector
\begin{equation}
\label{eq:fin1}
\vert \overline{\Psi}(\varepsilon,t)\rangle= \left(\begin{array}{l} \vert\Psi_I(t)\rangle  \\ \vert {\Psi}_R(t)\rangle\\  \vert {\Psi}_{A}^{(\sigma/\delta)} (\varepsilon,t) \rangle \\  \vert {\Psi}_{A}^{\pi} (\varepsilon,t) \rangle \end{array} \right),
\end{equation}
where we have split the contributions to the final ionic state CO$^+(A~^2\Pi)$ produced by the emission of  $\varepsilon (\sigma/\delta)$ electrons and those by  the emission of  $\varepsilon \pi$ electrons. We define the matrix
\begin{multline}
\label{eq:fin2}
\hat{\mathbf{H}}(R,\theta,t)= \hat{\textbf{T}}(R,\theta)+ \\ 
\left(\begin{array}{llll}V_I(R)-\frac{i}{2}\Gamma_{ph}(t) & \left(D^\dag_x(t)-\frac{i}{2} W^\dag(t)\right) \sin \theta &0&0\\
\left(D_x(t)-\frac{i}{2} W(t)\right) \sin \theta ~~ & V_R(R) -\frac{i}{2}[\Gamma_{aug}+\Gamma_{\ast}(t)]-\omega ~~~ &  0&0\\
 d_x(t)\sin \theta & V& V_{A}(R)+\varepsilon  -\omega &0\\
 d_z(t)\cos\theta  &0&0&V_{A}(R)+\varepsilon  -\omega  \end{array} \right).
\end{multline}
With these notations, the final set of equations describing propagation of the nuclear wave packets (\ref{eq:fin1})  now takes on  the following compact form
\begin{equation}
\label{eq:fin3}
i \vert \dot{\overline{\Psi}}(\varepsilon,t)\rangle= \hat{\mathbf{H}}(t) \,  \vert  {\overline{\Psi}}(\varepsilon,t)\rangle.
\end{equation}
The matrix $\hat{\mathbf{H}} $ can be viewed as the effective Hamiltonian governing the two-dimensional nuclear  dynamics in the considered RA decay process  in an intense laser field. Below we summarize the physical meaning of each term of the Hamiltonian matrix (\ref{eq:fin2}) and provide their explicit expressions.

The $\hat{\textbf{T}}(R,\theta)$ matrix is the common nuclear kinetic energy operator for the vibrational motion along the internuclear distance $R$ and the rotational motion described by the angle $\theta$ between the polarization vector of the laser pulse and the molecular axis. The functions  $V_I(R)$, $V_R(R)$, and $V_A(R)$ on the diagonal are the potential energies of the corresponding electronic states. The transitions between the electronic states  are given by the following matrix elements of the total Hamiltonian (\ref{eq:hamilt}) 
\begin{subequations}
\label{eq:couplings}
\begin{equation}
\label{eq:couplingsRexc}
\langle \Phi_R\vert \hat{H}(t) \vert \Phi_I\rangle= \langle 2\pi\vert \hat{x}\vert 2\sigma\rangle \frac{\mathcal{E}_0 \,g(t)}{2}\, e^{-i\omega t}= D_x(t)\, e^{-i\omega t},
\end{equation}
\begin{equation}
\label{eq:couplingsRAD}
\langle\Phi_{A}^{\varepsilon(\sigma/\delta)} \vert \hat{H}(t) \vert  \Phi_R \rangle= \langle2\sigma \,\varepsilon(\sigma/\delta)\vert {1}/{\hat{r}_{12}}\vert 1\pi\,2\pi\rangle= V,
\end{equation}
\begin{equation}
\label{eq:couplingsdirsin}
\langle\Phi_{A}^{\varepsilon(\sigma/\delta)} \vert \hat{H}(t) \vert  \Phi_I \rangle= 
\langle \varepsilon(\sigma/\delta) \vert \hat{x}\vert 1\pi\rangle \frac{\mathcal{E}_0 \,g(t)}{2}\, e^{-i\omega t}= d_x(t)\, e^{-i\omega t},
\end{equation}
\begin{equation}
\label{eq:couplingsdircos}
\langle\Phi_{A}^{\varepsilon\pi} \vert \hat{H}(t) \vert  \Phi_I \rangle= 
\langle \varepsilon\pi \vert \hat{z}\vert 1\pi\rangle \frac{\mathcal{E}_0 \,g(t)}{2}\, e^{-i\omega t}= d_z(t)\, e^{-i\omega t}.
\end{equation}
\end{subequations}
In Eqs.~(\ref{eq:couplingsRexc}), (\ref{eq:couplingsdirsin}), and (\ref{eq:couplingsdircos}), the   rotating wave approximation \cite{Griffiths94}  has already been utilized, and, in contrast to the rapidly oscillating factor $e^{-i\omega t}$, the functions  $D_x(t)$, $d_x(t)$ and $d_z(t)$ vary  slowly on the timescale of $2\pi/\omega$. The matrix element (\ref{eq:couplingsRexc}) describes the excitation of the resonant state via the $x$ component of the dipole transition operator. The matrix element (\ref{eq:couplingsRAD})  represents the Auger decay of the resonance into the final ionic state CO$^+(A~^2\Pi)$  with the emission of $\varepsilon(\sigma/\delta)$ electrons, including  both direct and exchange Coulomb integrals. The matrix elements (\ref{eq:couplingsdirsin}) and  (\ref{eq:couplingsdircos})  correspond to the direct ionization of the ground state into the final ionic state CO$^+(A~^2\Pi)$  with the emission of $\varepsilon(\sigma/\delta)$ or $\varepsilon\pi$  electrons, respectively, via corresponding components of the  dipole transition operator.  Here and below, all transition matrix elements are assumed to vary slowly with the energy across the resonance and with the geometry within the Franck-Condon region, and are replaced by their mean values.

The potential energy of the ground state $V_I(R)$ in the Hamiltonian matrix (\ref{eq:fin2})  is augmented by the imaginary time-dependent  term $-\frac{i}{2}\Gamma_{ph}(t)$ describing the  leakage from the ground state due to its direct photoionization into all possible final ionic state (referred hereafter as `\emph{GS-leakage}' \cite{Demekhin11SFatom}). Its explicit expression was obtained in  \cite{Demekhin11SFatom} in the local approximation \cite{Cederbaum81,Domcke91} and reads:
\begin{equation}
\label{eq:solut_gph}
\Gamma_{ph}(t)= 2\pi \sum _j  \vert d_{x/z}^j(t)\vert^2,
\end{equation}
where $d_{x/z}^j(t)$ are the dipole transition matrix elements for the direct ionization of the ground electronic state into all possible final ionic states numerated by superscript $j$, and can be computed similarly to Eqs.~(\ref{eq:couplingsdirsin}) and  (\ref{eq:couplingsdircos}). The total probability for the direct photoionization of the ground state (\ref{eq:solut_gph}) is identical to the quantity  $\gamma_{ph}(t)$ introduced in Ref.~\cite{Liu10}:
\begin{equation}
\label{eq:solut_gelm}
\Gamma_{ph}(t)=\gamma_{ph}(t)=\sigma^{tot}_{ph}I(t)/\omega,
\end{equation}
where $\sigma^{tot}_{ph}$ is the total direct photoionization cross section of the ground state at exciting-photon energy $\omega$, $I(t)$ is the field intensity (\ref{eq:intens}), and the quantity $I(t)/\omega$ stands in Eq.~(\ref{eq:solut_gelm}) for the photon flux.

Similarly, the total leakage from the resonance with `dressed' potential energy $V_R(R)-\omega$  is provided by the imaginary part $-\frac{i}{2}[\Gamma_{aug}+\Gamma_{\ast}(t)]$ on the respective diagonal element of $\hat{\mathbf{H}}$. The first term, $-\frac{i}{2} \Gamma_{aug}$, describes the usual time-independent leakage due to the Auger decay (`\emph{RA-leakage}' \cite{Demekhin11SFatom}). In the local approximation \cite{Cederbaum81,Domcke91}, the total  rate for the Auger decay is given by Coulomb matrix element similar to Eq.~(\ref{eq:couplingsRAD}) as  \cite{Pahl99ZPD}
\begin{equation}
\label{eq:tot_width}
\Gamma_{aug}= 2\pi  \sum _j \vert V_{j}\vert^2,
\end{equation}
where the summation over  index $j$ runs over all possible RA decay channels. The second term, $-\frac{i}{2}\Gamma_{\ast}(t)$, is the time-dependent leakage due to the direct photoionization of the resonance (`\emph{RD-leakage}' \cite{Demekhin11SFatom}). As a   good approximation \cite{Liu10,Sun10} it can be chosen equal to GS-leakage, $\Gamma_{\ast}(t)=\Gamma_{ph}(t)$, since only outer electrons participate in the direct ionization of the ground state and the resonance at chosen X-ray photon energy $\omega$.

As was demonstrated in Ref.~\cite{Demekhin11SFatom}, the usual direct coupling $D_x(t)$ between the ground state and the resonance through the laser field \cite{Rohringer08,Liu10,Sun10} is augmented by an additional  time-dependent term $-\frac{i}{2} W(t)$, named  Leakage-Induced Complex coupling (`\emph{LIC-coupling}').  This term appears only if the  photoionization from the ground state and   Auger decay are simultaneously treated as required. Explicitly  the LIC-coupling reads \cite{Demekhin11SFatom}
\begin{equation}
\label{eq:solut_LIC}
W(t)=2\pi \sum_j d^j_{x}(t)V^\dag_{j},
\end{equation}
where the summation over index $j$ runs over all possible final ionic states accessible by both channels. This   is an indirect coupling  which can be interpreted as follows: The photoelectron emitted by the ground state is recaptured by the residual ion to produce the resonance state,   and reversely, the Auger electron can be captured by the residual ion which then becomes the neutral atom in its ground state. The   ground  and `dressed' resonance states coupling (i.e., the matrix elements $\mathrm{H}_{12}$ and $\mathrm{H}_{21}$) is non-hermitian and   operative as long as the pulse is on.

We note that the matrix elements of the permanent dipole moment of the CO molecule in the ground and resonance states must also be included in the respective diagonals of the Hamiltonian matrix (\ref{eq:fin2}). However, the permanent dipole moment of CO in its ground state is small. In addition, these matrix elements are  proportional to the rapidly oscillating function $e^{-i\omega t}$, and the average effect of the permanent dipole moment on the pulse duration timescale is negligibly small. For simplicity, these matrix elements are not included in the present calculations.

The third row of the  Hamiltonian matrix (\ref{eq:fin2}) describes the nuclear dynamics on the studied  final ionic state  produced via the emission of   $\varepsilon(\sigma/\delta)$ electrons. The ionic state is populated  by the direct photoionization from the ground state (matrix element  $ d_{x}(t)\sin\theta $) and coherently by the Auger decay of the resonance (matrix element $ V  $), all at a given kinetic energy $\varepsilon$ of the emitted electron, and the created  wave packet propagates on the `dressed' PESs  $V_A(R)+ \varepsilon-\omega$. The same final ionic state can be produced by the direct ionization of the ground state via the emission of $\varepsilon\pi$ electron. This  process is governed  by the matrix element $ d_{z}(t) \cos\theta $ in the  forth row of the  Hamiltonian matrix (\ref{eq:fin2}). The matrix element  $\mathrm{H}_{42}=0$ indicates that the resonant population via the emission of $\varepsilon\pi$ electron is forbidden. Separating these two different kinds of processes into  two different lines of the Hamiltonian matrix ensures that the corresponding wave packets accumulate on the PES of the CO$^+(A~^2\Pi)$  ionic state incoherently (see also discussion above).

Finally, the nuclear wave packets $\vert {\Psi}_{A}^{(\sigma/\delta)} (\varepsilon,t) \rangle $ and $  \vert {\Psi}_{A}^{\pi} (\varepsilon,t) \rangle$ contain the information on the  RA electron spectrum for the production of the CO$^+(A~^2\Pi)$ ionic state. The spectrum can be computed as the incoherent  sum of the norms of the respective wave packets at  long times \cite{Pahl99ZPD}
\begin{equation}
\label{eq:spectrum}
\sigma_A(\varepsilon) = \lim_{t\to\infty}  \langle  {\Psi}_{A}^{(\sigma/\delta)} (\varepsilon,t)  \vert {\Psi}_{A}^{(\sigma/\delta)} (\varepsilon,t) \rangle +  \lim_{t\to\infty}  \langle  {\Psi}_{A}^{\pi} (\varepsilon,t)   \vert{\Psi}_{A}^{\pi} (\varepsilon,t) \rangle  .
\end{equation}

\begin{table}
\caption{Parameters for photoionization (PI) and resonant Auger (RA) decay of CO utilized in the present calculations. The energy of the exciting radiation  is chosen to be $\omega=287.4$~eV. }
\begin{ruledtabular}
\begin{tabular}{ll}
State/Quantity/Source& Value    \\\hline
\multicolumn{2}{l}{Ground state $1\sigma^2 2\sigma^2 3\sigma^2 4\sigma^2 1\pi^4 5\sigma^2 ~X~ ^1\Sigma^+$}  \\ 
Total direct PI cross section  \cite{Hemmers93,BeckShirl}& $\sigma_{ph}^{tot} =0.24$~Mb    \\\hline
\multicolumn{2}{l}{Resonance $1\sigma^2 2\sigma^1 3\sigma^2 4\sigma^2 1\pi^4 5\sigma^2 2\pi^1~ ^1\Pi$}  \\  
Total RA decay rate \cite{Prince99}& $\Gamma_{aug}=80$~meV   \\
Oscillator strength \cite{Demekhin09CstarO}& $f =0.158$~a.u.  \\
\hline
\multicolumn{2}{l}{Participator state  $1\sigma^2 2\sigma^2 3\sigma^2 4\sigma^2 1\pi^3 5\sigma^2 ~A~ ^2\Pi$  }  \\ 
Partial RA decay rates  \cite{Demekhin09CstarO}\\
$\varepsilon\sigma$ channel& $\Gamma_{aug}^{A\varepsilon\sigma}= 1.07$~meV   \\
$\varepsilon\delta$ channel& $\Gamma_{aug}^{A\varepsilon\delta}= 3.46$~meV   \\
Partial direct PI  cross sections  \cite{Demekhin09CstarO}\\
$\varepsilon\sigma$ channel & $\sigma^{A\varepsilon\sigma}_{ph} =0.0060$~Mb   \\
$\varepsilon\pi$ channel& $\sigma^{A\varepsilon\pi}_{ph} =0.0105$~Mb  \\
$\varepsilon\delta$ channel & $\sigma^{A\varepsilon\delta}_{ph} =0.0185$~Mb  \\
\end{tabular}
\label{tab:param}
\end{ruledtabular}
\end{table}

\subsection{Intersections of complex energy surfaces}
\label{sec:thyDICES}

\begin{figure}
\includegraphics[scale=0.55]{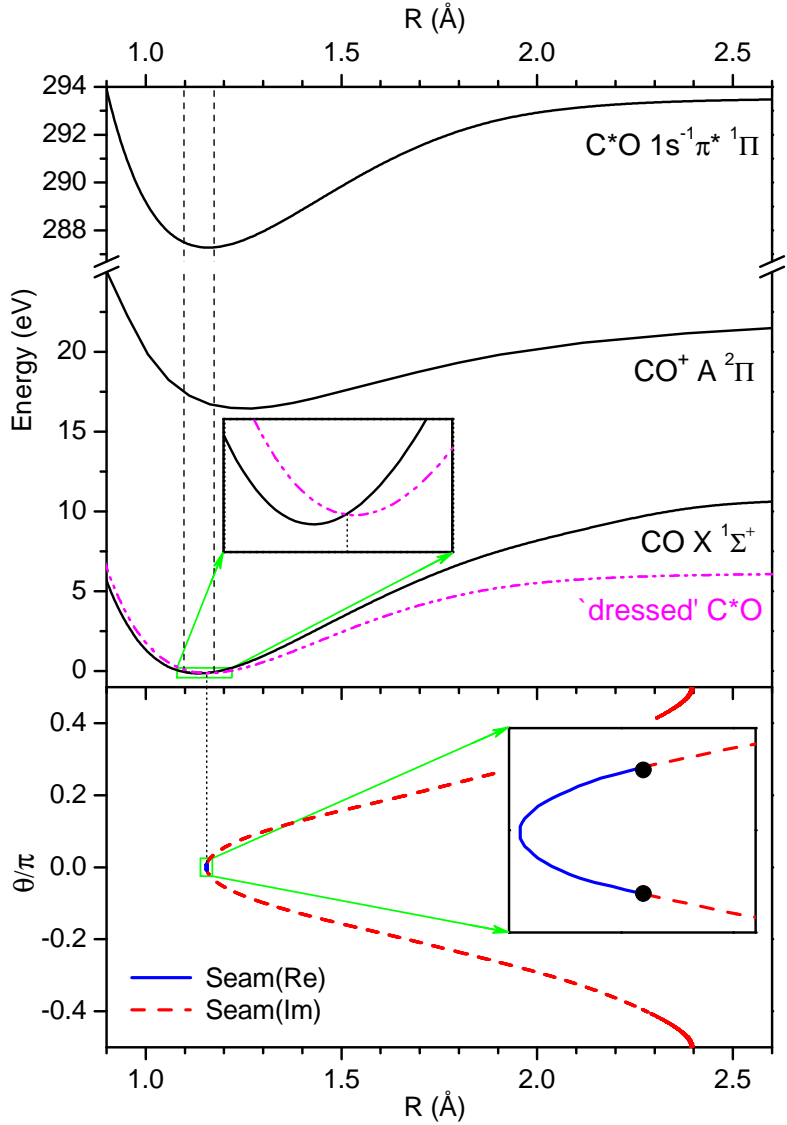}
\caption{(Color online) \emph{Upper panel:} The potential energy curves of the ground $X~^1\Sigma^+$ state of CO, the resonant state C$^\ast$O($1s^{-1}\pi^\ast~^1\Pi$), and the final ionic state CO$^+(A~^2\Pi$) from Ref.~\cite{Demekhin09CstarO}, together with the curve of the resonant state `dressed' by the photon energy $\omega=287.4$~eV. The Franck-Condon region for the ground state of CO is marked by the vertical dashed lines and the position of  curve crossing is indicated by vertical doted line (see also inset). \emph{Lower panel:} When the molecule is exposed to the X-ray laser field of peak intensity $I_0=3\times10^{16}$~W/cm$^{2}$, the ground and resonant  potentials become doubly intersecting complex surfaces in $R$ and $\theta$ space. Solid and dashed lines represent the  projections of the real and imaginary seams of the complex intersecting surfaces onto the $R\theta$ plane. Two solid circles in the inset indicate  the complex  intersection points, where the real surfaces and the imaginary surfaces are  simultaneously degenerate (see also Fig.~\ref{fig:imag}). }
\label{fig:pecs}
\end{figure}

As   mentioned in the introduction, all parameters needed for the present calculations of the nuclear dynamics governed by the  Hamiltonian matrix (\ref{eq:fin2}), such as the total and partial direct photoionization cross sections and the RA decay rates, as well as the oscillator strength for the resonant excitation, can be found in the literature. These parameters are collected in Tab.~\ref{tab:param}. The potential energy curves of the ground, the resonant, and chosen final ionic states computed in Ref.~\cite{Demekhin09CstarO} are depicted in the upper panel of Fig.~\ref{fig:pecs}. The  vertical excitation energy from the initial state to the  $v_r=0$ vibrational level  of the  C$^\ast$O(1s$^{-1}\pi^\ast$) resonance  is  $\omega=287.4$~eV \cite{Piancastelli97}. Of all levels, this level possesses  the maximal excitation probability which is  about 87\% \cite{Piancastelli97,Demekhin09CstarO}.  The energy curve of the resonant state `dressed' by the field of this energy (i.e., $V_R(R)-\omega$) is also depicted in the upper panel of Fig.~\ref{fig:pecs}. One can see, that this  curve   crosses the curve of the ground state in the  Franck-Condon region   (see also inset in this panel).

\begin{figure}
\includegraphics[scale=0.55]{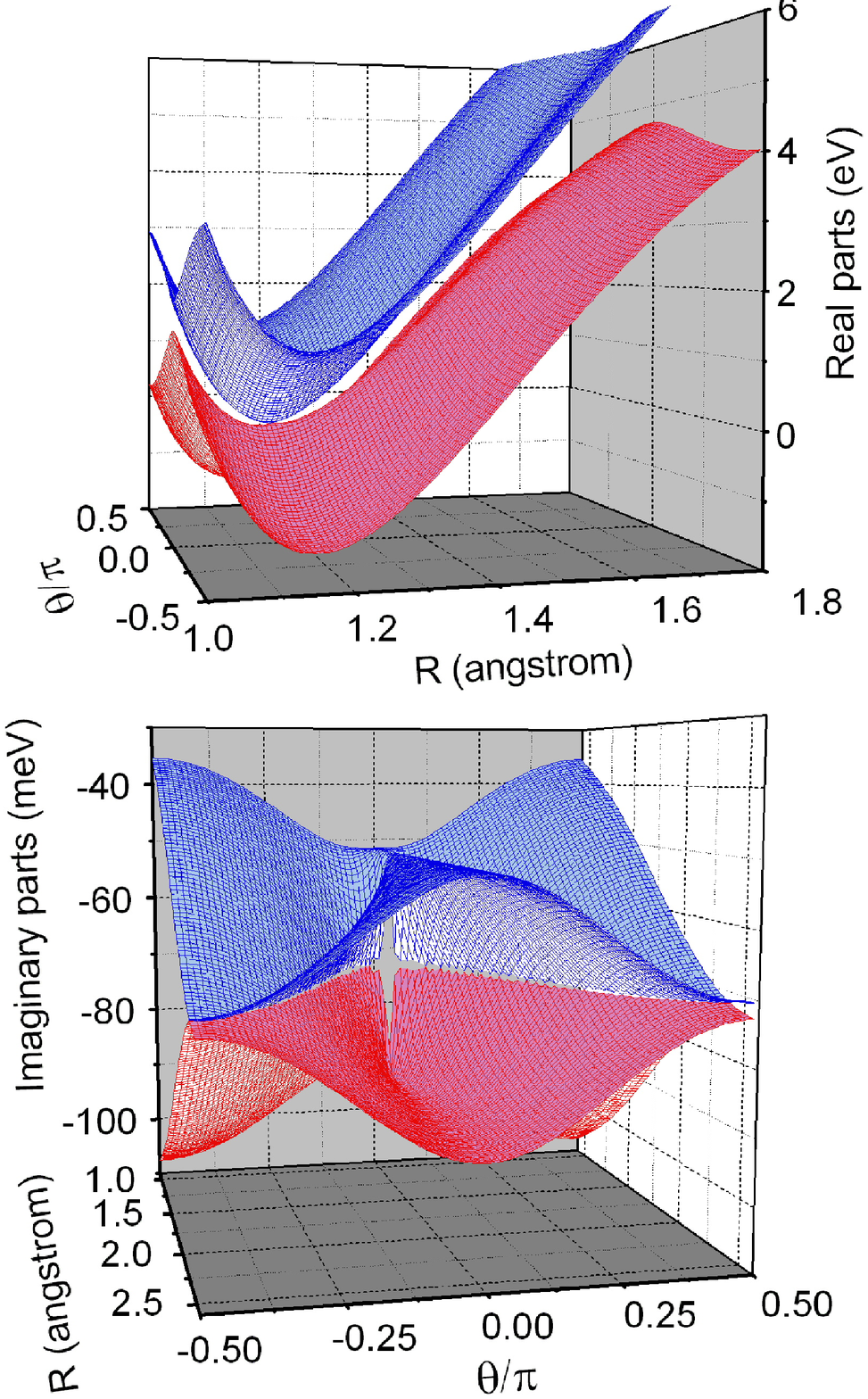}
\caption{(Color online) Real (upper panel) and imaginary (lower panel) parts of the resulting two-dimensional complex potential energy surfaces. The topology of the surfaces is that of doubly intersecting complex energy surfaces (DICES).}
\label{fig:imag}
\end{figure}

As has been demonstrated in  Ref.~\cite{MolRaSfPRL}, crossing of  `dressed'  by field  states of a diatomic molecule results in the  intersections of the corresponding two-dimensional   complex PESs. It is well known \cite{Domcke04,Baer} that for the formation of a conical intersection  one needs at least two nuclear degrees of freedom whose changes affect the electronic wave function. These are not available for a free diatomic molecule. However, two dynamical variables $R$ and $\theta$ enter  the Hamiltonian matrix  (\ref{eq:fin2})  explicitly and the rotation becomes the missing degree of freedom to allow for the formation of  intersections for diatomics in the presence of a laser field. We point out that the rotational degree of freedom is involved in the nuclear dynamics only due to the presence of the laser field and only when the pulse is on. Very important, the non-adiabatic couplings, i.e., matrix elements of the nuclear momenta along $R$ and $\theta$, between the two `dressed' electronic states are singular at these intersections \cite{DICES}. This gives rise to dramatic dynamical effects.

Without the on- and off-diagonal  imaginary corrections to the  Hamiltonian matrix  (\ref{eq:fin2}), the two-dimensional potential energy surfaces of the `dressed' electronic states  exhibit an intersection at  $\theta=0$ (coupling matrix elements $\mathrm{H}_{12}$ and $\mathrm{H}_{21}$ are proportional to $\sin\theta$). The impact of such light-induced conical intersections on the dynamics of a system has been recently demonstrated in Ref.~\cite{Sindelka11}. Due to the presence of RA decay width, leakages and LIC-coupling in (\ref{eq:fin2}), the situation becomes more complicated. The two potential energy surfaces in $R$ and $\theta$  space  obtained by diagonalizing the electronic Hamiltonian  $\hat{\mathbf{H}}(R,\theta,t)- \hat{\textbf{T}}(R,\theta) $ in Eq.~(\ref{eq:fin2}) are now complex and generally exhibit two intersecting points at  which  the real as well as the imaginary parts of the two electronic energies become degenerate \cite{DICES}. This analogue of a conical  intersection in the continuum has been named   doubly intersecting complex energy surfaces (DICES).

In order to illustrate the DICES appearing in the present case we have diagonalized  the complex  electronic Hamiltonian. The `dressed'  potential energy curves shown in the upper panel of  Fig.~\ref{fig:pecs} and  parameters relevant for the RA decay of CO listed in Tab.~\ref{tab:param} were utilized  for the peak intensity of the  laser pulse of $I_0=3\times10^{16}$~W/cm$^{2}$. The real and imaginary parts of the resulting surfaces are shown in Fig.~\ref{fig:imag}. The projections of the real and imaginary seams of the complex intersecting surfaces onto the $R\theta$ plane are shown in the lower panel of  Fig.~\ref{fig:pecs}.  The seam of degeneracy of the real parts of the surfaces is shown  by the solid line (see also inset for an enlarged scale). The imaginary parts of the surfaces have a complementary seam of degeneracy shown  by the dashed line. The complex surfaces intersect  at exactly two points which are at the edges of the seams (solid circles in the inset). The present case fits well to the discussion of DICES considered in  Ref.~\cite{DICES}. Finally, we note that the imaginary parts of the surfaces which reflect the decay of the resulting `dressed' states are now strongly dependent on $R$ and $\theta$ \cite{MolRaSfPRL,DICES} (see the lower panel of Fig.~\ref{fig:imag}) in contrast to the initially constant values entering  the Hamiltonian matrix (\ref{eq:fin2}).

\section{Results and discussion}
\label{sec:res}

In order to illustrate the impact of DICES on the dynamics of the RA decay we performed three sets of calculations to which we refer as models. The first model is the exact calculation including both vibrational and rotational degrees of freedom in the nuclear dynamics (full Hamiltonian matrix (\ref{eq:fin2})). We refer to this model as the `\emph{DICES}' model. In the second model, only the vibrational motion is allowed in the nuclear dynamics, eliminating thereby DICES (hereafter referred  to as the `\emph{no DICES}' model). To exclude rotations from the Hamiltonian matrix (\ref{eq:fin2}), we apply the common approach  \cite{Sindelka11,optic1,optic2,optic3}  incorporating rotational transitions via the selection rule $\Delta J = \pm 1$. The approximate Hamiltonian used for the vibrational dynamics of diatomics in laser fields can be obtained in this approach by replacing $\cos\theta$ and  $\sin \theta$  functions in Eq.~(\ref{eq:fin2}) by their optical transition matrix elements $\langle Y_{10}\vert \cos\theta \vert Y_{00}\rangle =\sqrt\frac{1}{3}$ and $\langle Y_{1\pm 1}\vert \sin\theta \vert Y_{00}\rangle=\sqrt\frac{2}{3}$, respectively, where $Y_{lm} $ are the usual spherical harmonics. In the third model (hereafter referred to as the `\emph{Atomic}' model), the whole nuclear dynamics is excluded from the calculations. For this purpose, the internuclear distance $R$ is kept fixed at the position of the curves crossing (see Fig.~\ref{fig:pecs}) and $\cos\theta$($\sin \theta)$ in Eq.~(\ref{eq:fin2}) are replaced by their optical transition matrix elements as in the  \emph{no DICES} model.

Within all three models, we also study the individual contributions of different leakage mechanisms to the RA effect. For this purpose, we performed the full calculations as well as calculations in which we take into account only one or two of the leakage mechanisms. We shall address the results of these systematic approximations to the three models as:
\begin{itemize}
\item `\emph{Resonant}' -- only the resonant channel is accounted for, i.e., all losses due to photoionization are neglected;
\item  `\emph{Direct}'  -- only the direct ionization channel is taken into account, i.e., the resonance state is excluded from the calculations; 
\item `\emph{Interference}' --  both the direct and resonant channels and the interference between them are taken into account, but the resonance state itself cannot  photoionize;
\item `\emph{Total}' --  all mechanisms are taken into account including the direct photoionization of the resonance.
\end{itemize}
The present calculations were performed for Gaussian-shaped laser pulses of durations $\tau=4,~8$, and 16~fs which are comparable to the RA decay lifetime of the  C$^\ast$O resonance of $\tau_d\approx 8.2$~fs. To be able to carry out the two-dimensional calculations on the coupled complex surfaces we employed an efficient  Multi-Configuration Time--Dependent Hartree (MCTDH) method \cite{Meyer90mctdh} and code \cite{MCTDH}.

In order to evaluate the LIC-coupling $W(t)$, one has to include the manifold of all final ionic states populated coherently via resonant and direct ionization channels (sum over index $j$ in Eq.~(\ref{eq:solut_LIC})). These individual contributions of different final states to $W(t)$ can be different even by sign. In the absence of necessary data on all final ionic states we have estimated and utilized in the calculations an upper bound of the true value of $W(t)$. For this purpose we assumed that only one direct dipole transition amplitude $d_x(t)$ and only one Auger decay Coulomb matrix element $V$ enter Eq.~(\ref{eq:solut_LIC}), which, however, correspond to the total direct photoionization cross section and to the total RA decay probability (Tab.~\ref{tab:param}), respectively. Our computations shown below illustrate that for CO the impact of the LIC-coupling is moderate for the field intensities investigated.

\subsection{Total electron yield}
\label{sec:Yield}

\begin{figure}
\includegraphics[scale=0.6]{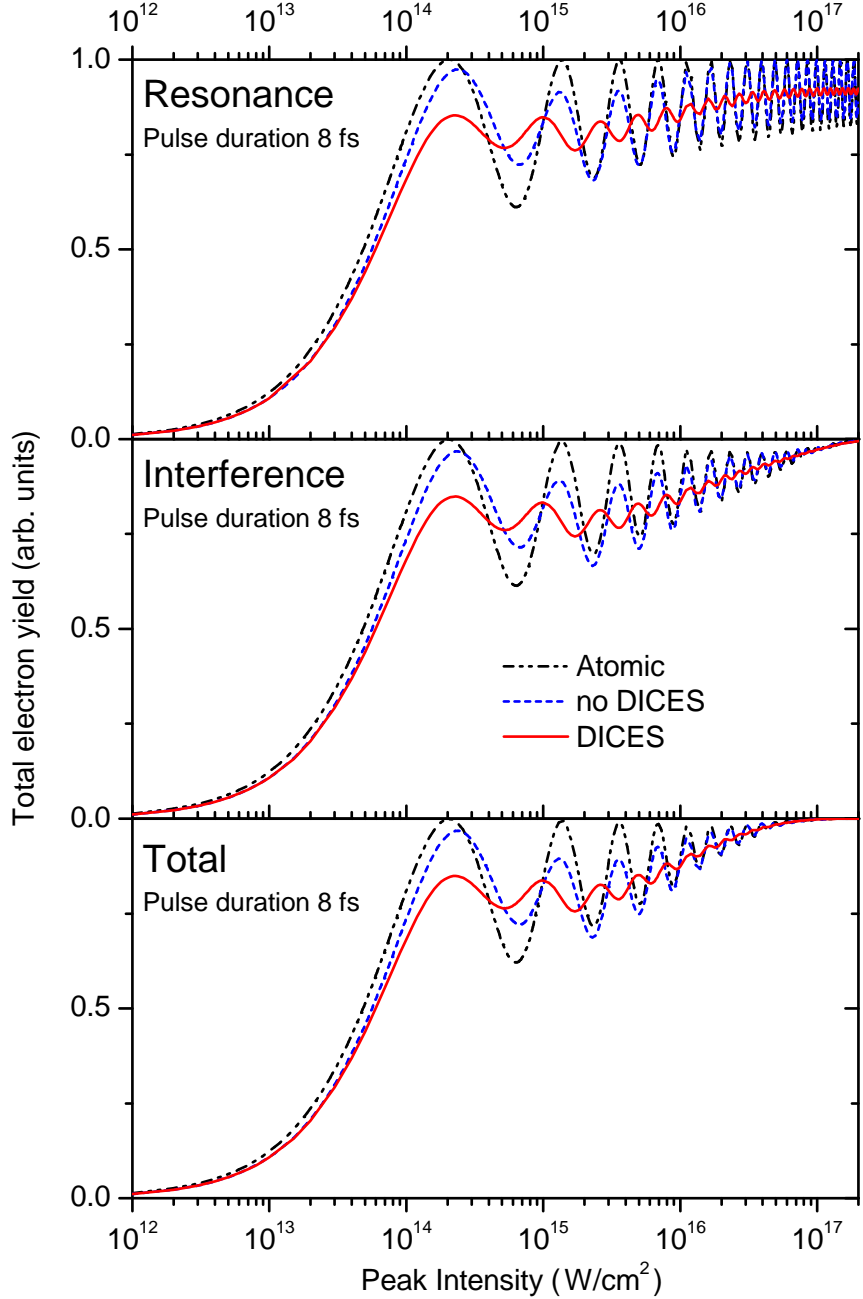}
\caption{(Color online) Total electron yield after exposure of CO to a coherent Gaussian-shaped pulse of duration  8~fs computed within different models and approximations discussed in the text. These include the atomic-like model (\emph{Atomic}; dash-dotted curves), rotations free model (\emph{no DICES}; dashed curves), and the exact model accounting   for  the vibrational and rotational degrees of freedom (\emph{DICES}; solid curves). \emph{Upper panel:} contribution of only the resonant channel  (\emph{Resonant}). \emph{Middle panel:} contribution  of both the direct and resonant channels and the interference between them (\emph{Interference}).  \emph{Lower panel:} the contributions of all mechanisms  including the direct photoionization of the resonance (\emph{Total}).}
\label{fig:yield_models}
\end{figure}

\begin{figure}
\includegraphics[scale=0.6]{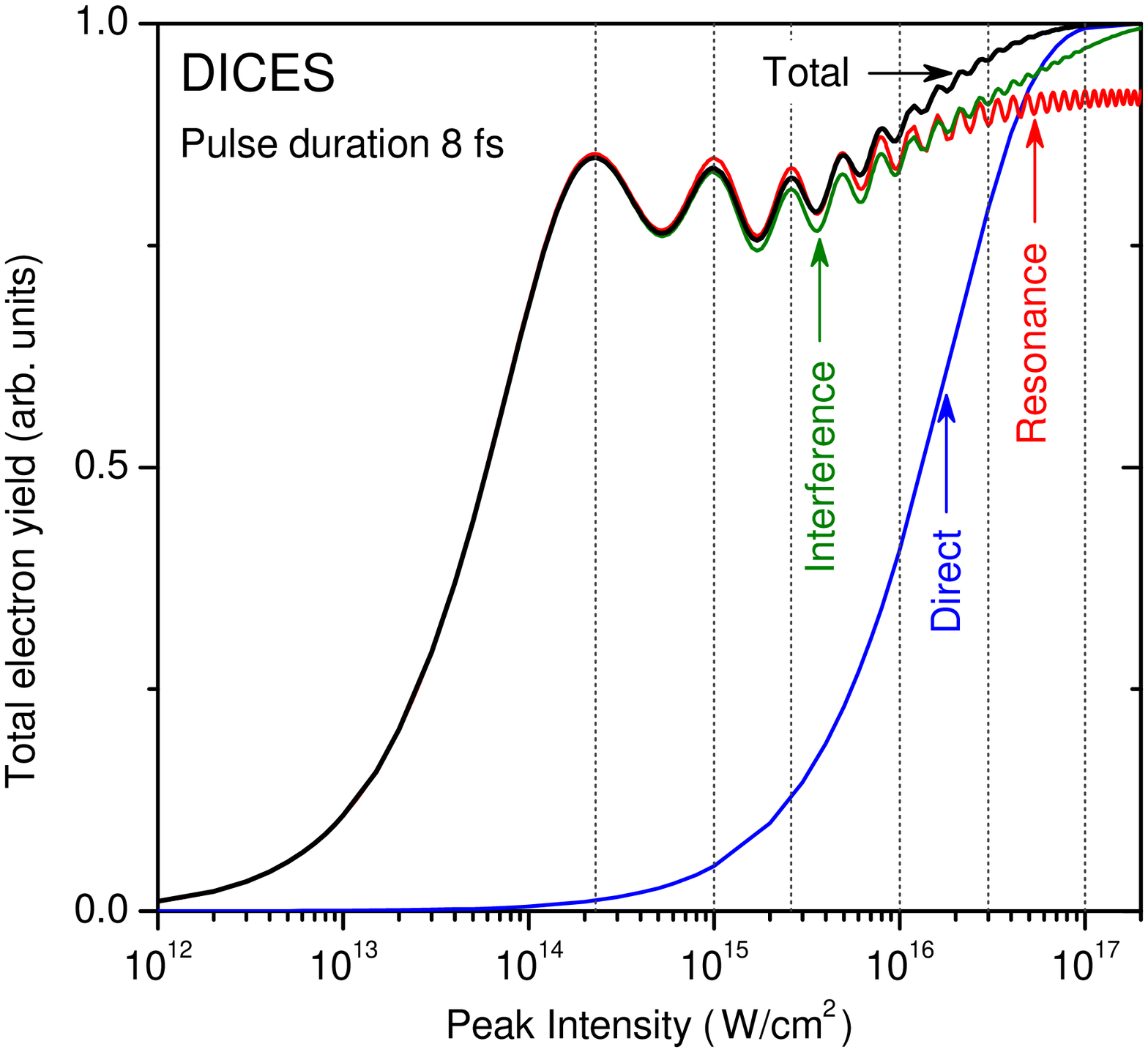}
\caption{(Color online) Total electron yield after exposure of CO to a coherent Gaussian-shaped pulse of duration  8~fs computed within the exact model, accounting   for  the vibrational and rotational degrees of freedom (\emph{DICES} model). Shown are   results of  different approximations  discussed in the text. These include the contribution of only the resonant channel  (\emph{Resonant}), of only the direct ionization channel (\emph{Direct}), and of both the direct and resonant channels and the interference between them (\emph{Interference}), as well as the contributions of all mechanisms  including the direct photoionization of the resonance (\emph{Total}). The vertical dotted lines indicate the peak intensities chosen for the calculation of the Auger spectra in Figs.~\ref{fig:spect_8fs1} and \ref{fig:spect_8fs2}.}
\label{fig:yield_approx}
\end{figure}

\begin{figure}
\includegraphics[scale=0.6]{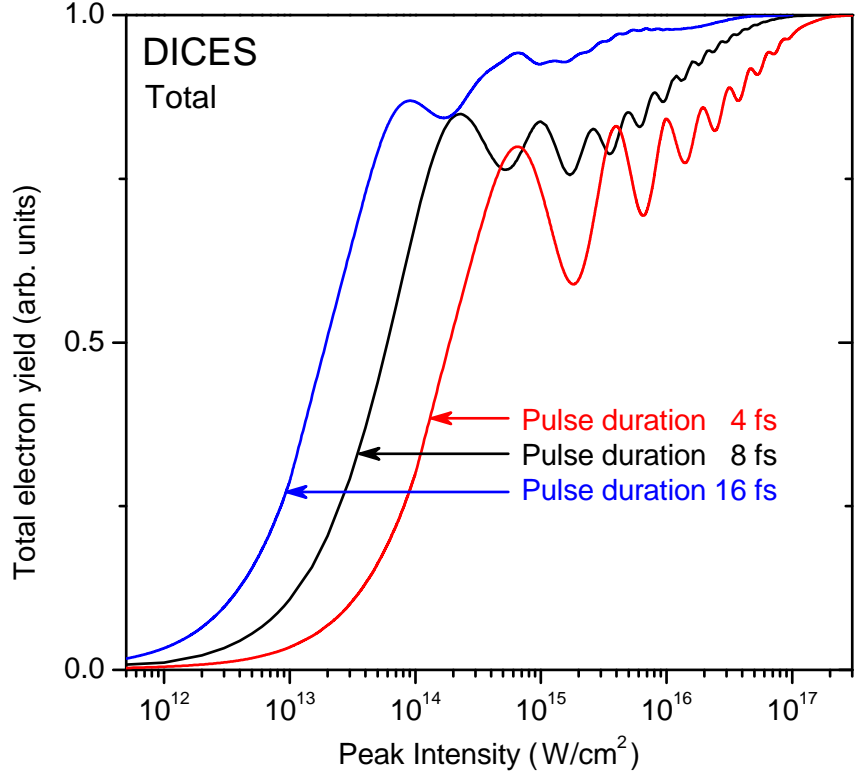}
\caption{(Color online) Total electron yield after exposure of CO to  coherent Gaussian-shaped pulses of different  durations (4, 8, and 16~fs) computed within the \emph{DICES}   model which accounts   for  the vibrational and rotational dynamics, employing the full formalism  (\emph{Total}: no approximation).}
\label{fig:yield_pulse}
\end{figure}

The total electron yield as a function of the X-ray peak intensity calculated in the different models and approximations discussed above is depicted in  Figs.~\ref{fig:yield_models}, \ref{fig:yield_approx},  and \ref{fig:yield_pulse}. In Fig.~\ref{fig:yield_models}, the impact of DICES and nuclear dynamics in the presence of different leakage mechanisms is illustrated for a pulse duration of 8~fs. Fig.~\ref{fig:yield_approx} illustrates the impact of different leakage mechanisms within the DICES model for the same pulse duration. The total electron yield computed exactly (i.e., within the \emph{DICES} model without approximations (\emph{Total})) is compared in Fig.~\ref{fig:yield_pulse} for different pulse durations. The total electron yield was computed as in  \cite{Rohringer08,Demekhin11SFatom} using
\begin{equation}
\label{eq:yield}
\lim_{t\to\infty} \sum_j\int d\varepsilon_j  \langle  {\Psi}_j (\varepsilon_j,t)   \vert {\Psi}_j (\varepsilon_j,t) \rangle     = 1-\lim_{t\to\infty} \langle  {\Psi}_{I} (t)   \vert{\Psi}_{I} (t) \rangle ,
\end{equation}
where index $j$ in the left-hand side of Eq.~(\ref{eq:yield}) must run over all possible final ionic states.

The atomic-like case, where the nuclear dynamics in $R$ and $\theta$ space is excluded completely, is well understood \cite{Rohringer08,Demekhin11SFatom}.  In the \emph{Resonant} approximation within the \emph{Atomic} model, only the Auger decay of the resonance and the Rabi oscillations between the resonance and the ground state are competing with each other. If the duration of the pulse is comparable or shorter than the Auger decay lifetime of the resonance and its intensity is low, the system has a finite probability of staying neutral after the laser pulse is over (dashed-dotted curve in the uppermost panel of Fig.~\ref{fig:yield_models}; intensities below $10^{14}$~W/cm$^{2}$). At larger intensities the field manages  to transfer the whole population from the ground state into the resonance, and the atomic-like system can be completely ionized with the probability of 1. This situation corresponds to the half-completed Rabi cycles and to maxima in the total electron yield arriving at unity \cite{Rohringer08} (dashed-dotted curve in the uppermost panel of Fig.~\ref{fig:yield_models}). At certain intensities the atomic system manages to complete several Rabi cycles during the pulse (to transfer the population back to the ground state) and the ionization probability drops again. These intensities correspond to the minima in the total yield.

In the \emph{Resonance} approximation for a molecule, the nuclear motion is additionally involved in the dynamics of the RA decay. As a result, the oscillations in the total electron yield possess much smaller amplitudes. Even in the \emph{no DICES} model, where only $R$ is the dynamical variable and $\theta$ is excluded from the dynamics, the difference from the \emph{Atomic} model is substantial \cite{MolRaSfPRL}  (cf dash-dotted and dashed curves in the uppermost panel of Fig.~\ref{fig:yield_models}). The maxima in the total electron yield are not necessarily touch unity (dashed curve). This is because a part of the energy of the absorbed photons is now transferred from the electronic degrees of freedom to the nuclear vibrational motion. The deviation from the atomic case is even more dramatic if we compare with the results of the full dynamics involving $R$ and $\theta$  motions (\emph{DICES} model). In the \emph{Resonant} approximation within this model, the complexity of the energy surfaces stems from the RA-leakage mechanism (Auger decay width) alone, and the appearing DICES are somewhat different \cite{MolRaSfPRL} from the full case illustrated in the lower panel of Fig.~\ref{fig:pecs}. The seams of degeneracy of the real and imaginary parts of the surfaces belong to the  straight line in the $R\theta$ plane, but not to the  contour (see also the original Ref.~\cite{DICES}). The strong non-adiabatic effects produced by the DICES manage to transfer energy from the vibrational to the rotational degree of freedom of a molecule. Consequently, the oscillations in the total yield are strongly suppressed from below and from above, and the yield is always well below 1 (solid curve in the uppermost panel of Fig.~\ref{fig:yield_models}). 
We emphasize that the strong impact of rotations illustrated in Fig.~\ref{fig:yield_models} is not due to the rotational motion alone, but rather owing to the light-induced conical intersection (more precisely, DICES in the present case) which couples the vibrational and rotational motions \cite{Moiseyev08,Sindelka11,MolRaSfPRL}.  In both \emph{Atomic} and  \emph{no DICES} models, the coupling matrix elements $\mathrm{H}_{12}$ and $\mathrm{H}_{21}$ in  Eq.~(\ref{eq:fin2})   carry the factor $\sqrt\frac{2}{3}$ after averaging over the rotational motion (i.e. by the optical transition matrix elements of  $\sin\theta$). Therefore, periods of the oscillations in the total yield computed in these two models cover larger intervals of the field strength than in the \emph{DICES} model.

The total electron yield computed in the \emph{Direct} approximation within the \emph{DICES} model (blue curve in Fig.~\ref{fig:yield_approx}) illustrates the individual contribution of the GS-leakage mechanism to the ionization process of the molecule. The total cross section for the ionization of the outer shells of CO ($\sigma_{ph}^{tot} =0.24$~Mb, Tab.~\ref{tab:param}) is much smaller than the probability for the ionization via the resonant channel. Thus, the total electron yield saturates to 1 due to the GS-leakage at much larger intensities (at about $10^{17}$~W/cm$^{2}$) than in the \emph{Resonant} approximation. After its saturation, the direct ionization channel becomes comparable with the resonant one \cite{Demekhin11SFatom}.

In the \emph{Interference} approximation  within the \emph{DICES} model (middle panel of Fig.~\ref{fig:yield_models}; and also green line in Fig.~\ref{fig:yield_approx}), four competitive mechanisms are present in the ionization of the molecule. These are the RA-leakage and   GS-leakage discussed above,  the interference between these two channels,  and, of course, the wave packet nuclear dynamics on the strongly coupled PESs. The imaginary parts of the resulting complex intersecting energy surfaces are responsible for the leakages of the propagating wave packets. As is evident from the lower panel of Fig.~\ref{fig:imag}, the resulting complex energy surfaces are strongly dependent on the dynamical variables and on all leakages. Consequently, a strict separation between the ionization mechanisms is impossible in the adiabatic picture. A simplified picture can, however, be suggested in the diabatic representation (before diagonalization of the electronic part of the Hamiltonian matrix  (\ref{eq:fin2})). The nuclear wave packet propagates on the coupled complex surfaces of the ground and `dressed' resonant states. The parts of the wave packet, propagating on the resonant surface, decay into the final ionic state via the RA-leakage mechanism. The parts of the wave packet, which propagate on the ground states surface, decay via the GS-leakage mechanism, i.e., direct photoionization. The competition between these two leakage mechanisms results in distinct modifications of the yield in all three models (middle panel of Fig.~\ref{fig:yield_models}). The oscillations become less pronounced and saturate along the trend imposed by the \emph{Direct} approximation (cf also green and red curves in Fig.~\ref{fig:yield_approx}). In the \emph{Atomic} model, not all of the oscillations arrive now at unity yield (dash-dotted curve in the middle panel of Fig.~\ref{fig:yield_models}). This weak effect is a consequence of the complex coupling term (LIC-coupling) \cite{Demekhin11SFatom}. The dramatic differences between the computed total yields shown in the middle panel of  Fig.~\ref{fig:yield_models} illustrate the impact of the strongly coupled ro-vibrational motion via the DICES underlying the processes included in the \emph{Interference} approximation. The differences between the dash-dotted and dashed curves are due to the vibrational motion, and those between the dashed and solid curves -- due to the rotational motion and DICES.

The photoionization of the resonance included in the RD-leakage mechanism enhances the ionization of the molecule (\emph{Total} calculations in the lowermost panel of Fig.~\ref{fig:yield_models}; and also black curve in Fig.~\ref{fig:yield_approx}). It results in further losses of the wave packet propagating on the resonant surface, reducing its norm by the direct ionization of the resonance into preferably highly-excited final ionic states. The oscillations in the total electron yield now become even less pronounced and saturate much earlier as a function of the laser intensity (cf green and black curves in Fig.~\ref{fig:yield_approx}). We note that depending on the pulse duration and the parameters of the system, there is a certain interval of intensities where both the GS-leakage and RD-leakage mechanisms are comparatively weak \cite{Demekhin11SFatom}. For the CO molecule exposed to a Gaussian pulse of 8~fs, these are the intensities up to about $10^{16}$~W/cm$^{2}$ (see Fig.~\ref{fig:yield_approx}). The total electron yields computed exactly (i.e., in the \emph{DICES} model and without approximations (\emph{Total})) for the three pulse durations of 4, 8, and 16~fs are compared in Fig.~\ref{fig:yield_pulse}. For the shorter pulse of 4~fs, all leakages are weak up to about an intensity of $2\times 10^{16}$~W/cm$^{2}$, and for the longer pulse of 16~fs up to about an intensity of $5\times 10^{15}$~W/cm$^{2}$ which are twice larger and twice lower than the corresponding value of the 8~fs pulse, respectively. The dramatic effects of the DICES on the ro-vibrational dynamics persists also in the full calculations (cf, all curves in the lowermost panel of Fig.~\ref{fig:yield_models}). The collective impact of all leakage mechanisms and of DICES results in a strong suppression of the oscillations of the total electron yield and in the saturation of the yield as a function of intensity, as illustrated in Fig.~\ref{fig:yield_approx} by the black curve.

\subsection{Resonant Auger spectrum}
\label{sec:spectra}

\begin{figure}
\includegraphics[scale=0.6]{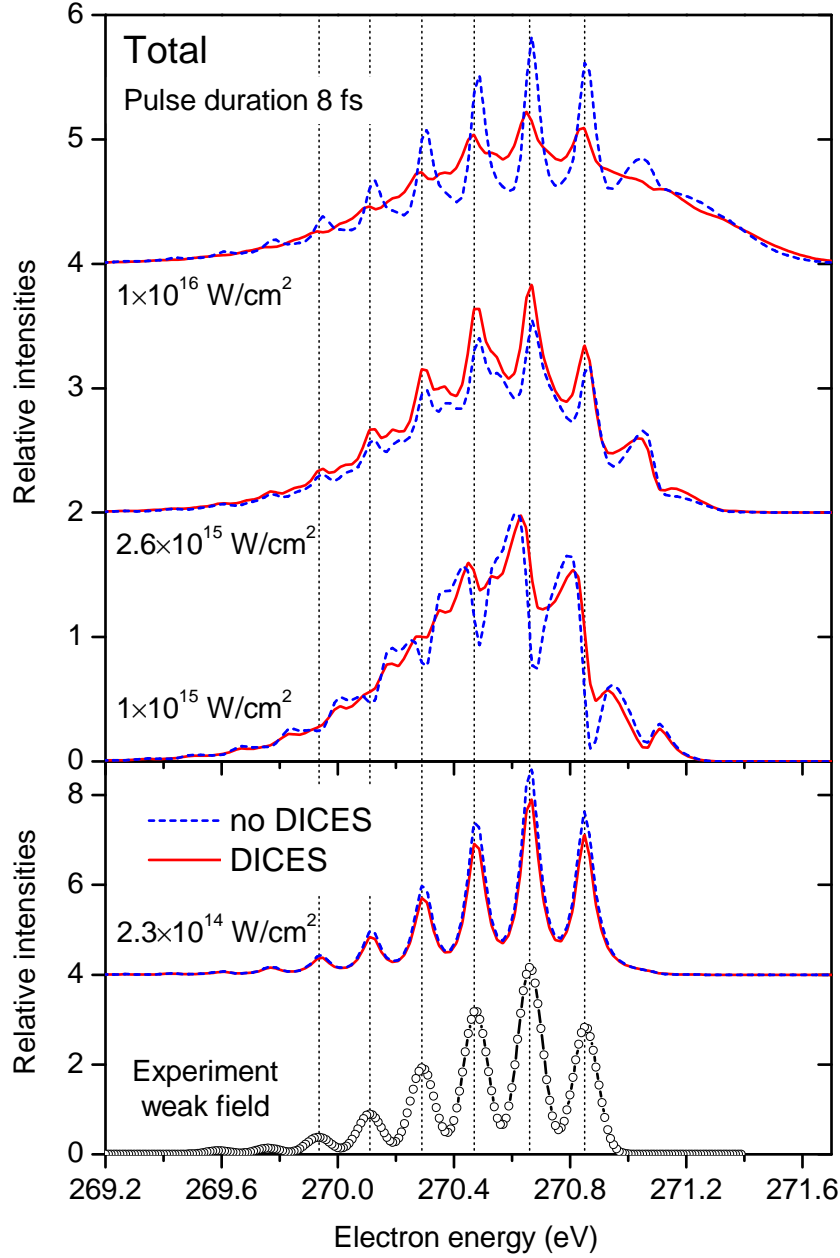}
\caption{(Color online) The RA electron spectra of CO  for the Gaussian-shaped pulse of 8~fs duration computed employing the full formalism (\emph{Total}) within the two models: \emph{no DICES} (broken curves) and \emph{DICES} (solid curves). The peak intensities $2.3\times 10^{14}$~W/cm$^{2}$, $1\times 10^{15}$~W/cm$^{2}$,   $2.6\times 10^{15}$~W/cm$^{2}$, and  $1\times 10^{16}$~W/cm$^{2}$ chosen for the calculations are marked  by the vertical dotted lines  in    Fig.~\ref{fig:yield_approx}.   The weak field experimental spectrum taken from Ref.~\cite{Kukk99} is shown for reference by open circles at the bottom.  The electron energies corresponding to the energy positions  of the $v^\prime $ vibrational levels of the CO$^+(A~^2\Pi$) state are marked by vertical dotted lines.}
\label{fig:spect_8fs1}
\end{figure}

\begin{figure}
\includegraphics[scale=0.6]{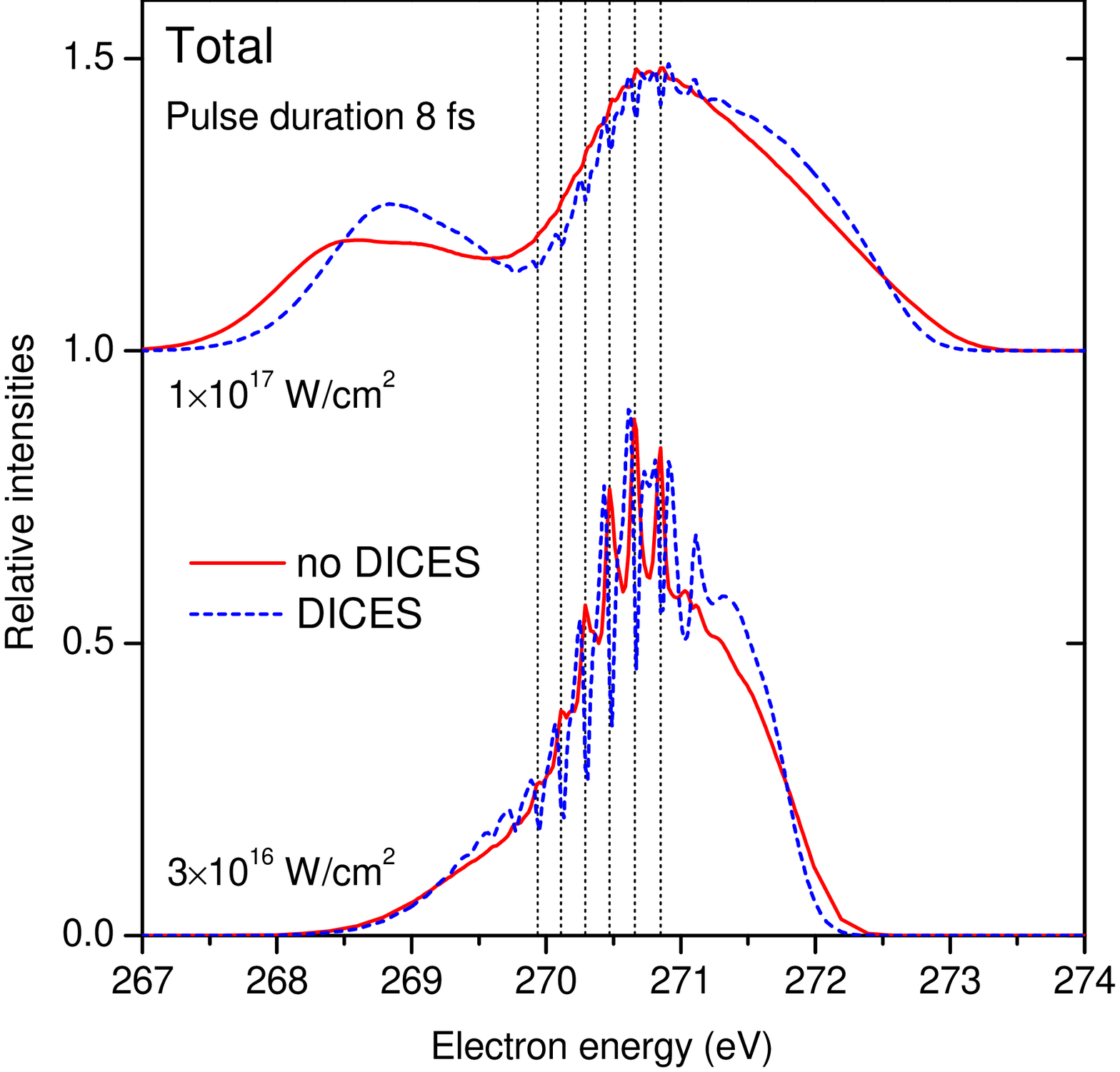}
\caption{(Color online) The RA electron spectra of CO  for the Gaussian-shaped pulse of   8~fs  duration computed  employing the full formalism (\emph{Total}) within the two models: \emph{no DICES} (broken curves) and \emph{DICES} (solid curves). The peak intensities are $3\times 10^{16}$~W/cm$^{2}$, and $1\times 10^{17}$~W/cm$^{2}$ (see also  notations in  Fig.~\ref{fig:spect_8fs1}).}
\label{fig:spect_8fs2}
\end{figure}

\begin{figure}
\includegraphics[scale=0.6]{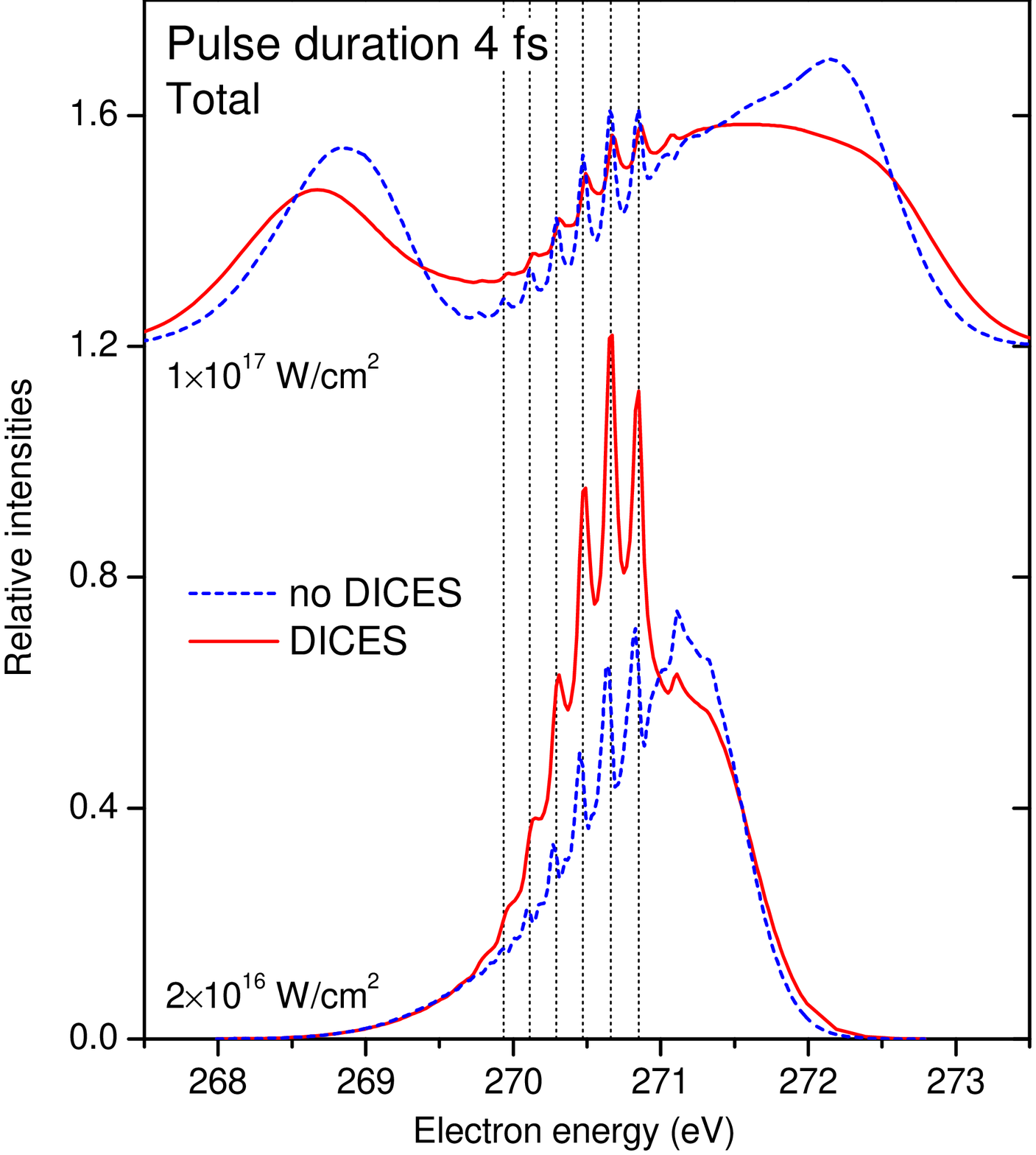}
\caption{(Color online) The RA electron spectra of CO  for the Gaussian-shaped pulse of 4~fs  duration computed  employing the full formalism (\emph{Total}) within the two models: \emph{no DICES} (broken curves) and \emph{DICES} (solid curves). The peak intensities are $2\times 10^{16}$~W/cm$^{2}$  and $1\times 10^{17}$~W/cm$^{2}$.}
\label{fig:spect_4fs}
\end{figure}

\begin{figure}
\includegraphics[scale=0.6]{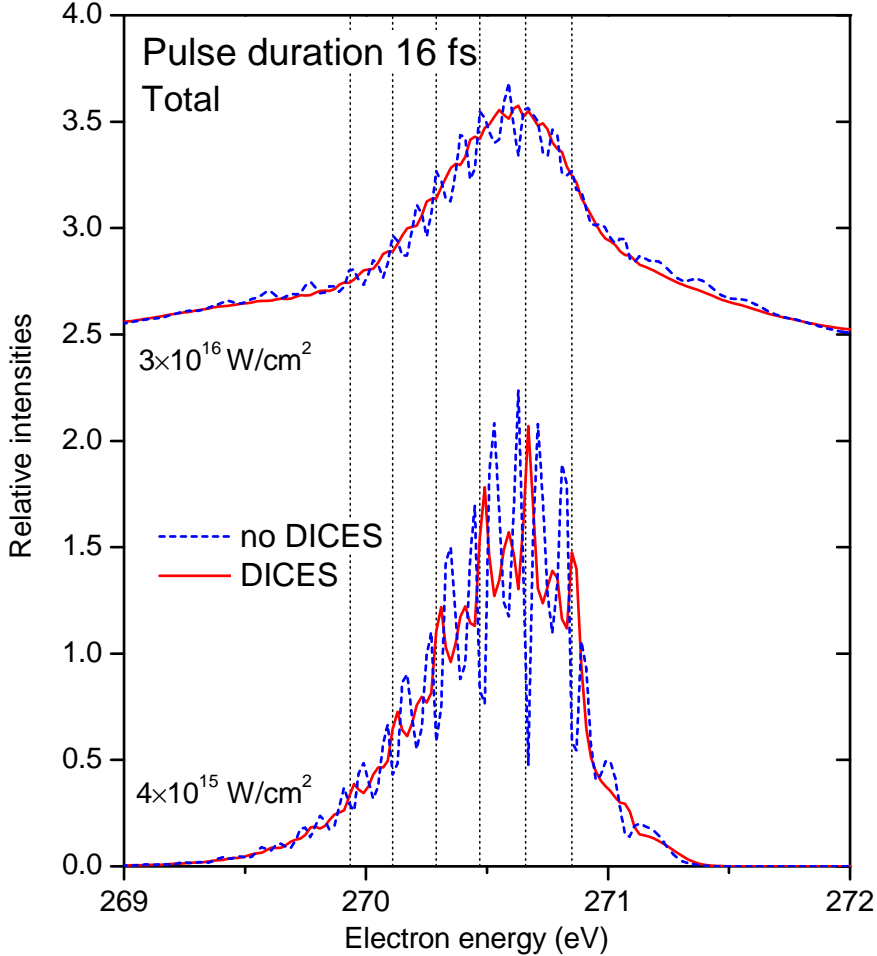}
\caption{(Color online) The RA electron spectra of CO  for the Gaussian-shaped pulse of   16~fs  duration computed  employing the full formalism (\emph{Total}) within the two models: \emph{no DICES} (broken curves) and \emph{DICES} (solid curves). The peak intensities are $4\times 10^{15}$~W/cm$^{2}$  and $3\times 10^{16}$~W/cm$^{2}$.}
\label{fig:spect_16fs}
\end{figure}

The RA electron spectra of the CO$^+(A~^2\Pi$) final ionic state at different peak intensities of the Gaussian-shaped pulse of 8~fs are depicted in Figs.~\ref{fig:spect_8fs1}  and \ref{fig:spect_8fs2}, and those for the shorter (4~fs) and longer (16~fs) pulses in Figs.~\ref{fig:spect_4fs} and \ref{fig:spect_16fs}, respectively. Shown in these figures are the results of calculations performed within the two models (\emph{no DICES} and \emph{DICES}) without further approximations (\emph{Total}), i.e., with all leakage mechanisms and interference effects included. The experimental spectrum of Ref.~\cite{Kukk99} measured in the weak field utilized by synchrotron radiation is shown for reference at the bottom of Fig.~\ref{fig:spect_8fs1} by open circles. One can see from the lower panel of Fig.~\ref{fig:spect_8fs1} that in a weak field (see the two spectra computed within the \emph{no DICES} and \emph{DICES} models and peak intensity of $2.3\times 10^{14}$~W/cm$^{2}$): (i) the structure of the spectrum is independent of whether we have included DICES or not; (ii) the contribution of all leakage mechanisms due to direct photoionization is negligible; and (iii) the present theory well reproduces the experiment \cite{Kukk99}.

Let us first discuss the effect of DICES on the RA electron spectra in the presence of all leakages, as well as the trends in the RA spectrum induced by the field intensity and pulse duration. These are illustrated in Figs.~\ref{fig:spect_8fs1}--\ref{fig:spect_16fs}.  At field intensities beyond the first maximum in the total electron yield (Fig.~\ref{fig:yield_approx}), the stimulated emission from the resonance starts to compete with the RA decay, resulting in significant modifications of the electron line profiles. In atoms the respective modifications consist in bifurcations of the RA electron peaks \cite{Rohringer08,Demekhin11SFatom}. However, the situation in molecules is more complicated than in atoms due to the presence of many vibrational and rotational levels of each electronic state involved in the process. At an intensity of $1\times 10^{15}$~W/cm$^{2}$ (lowermost spectra in the upper panel of Fig.~\ref{fig:spect_8fs1}), the local minima in the line shapes appear at energies where maxima of the intensity were observed in the weak field experiments, i.e., around  the electron energies corresponding to the energy positions of the $v'$ vibrational levels of the final ionic state marked by the vertical dotted lines in the figure. At a peak intensity of $2.6\times 10^{15}$~W/cm$^{2}$ (middle spectra in the upper panel of Fig.~\ref{fig:spect_8fs1}), the line shapes bifurcate again and their maxima again appear at their weak field energies. However, relatively weak local maxima are now present in between the large maxima. Up to the peak intensity of $1\times 10^{16}$~W/cm$^{2}$ shown in Fig.~\ref{fig:spect_8fs1}, the effect of DICES is strongly visible, but there are no truly dramatic effects to be seen when comparing the \emph{no DICES} and \emph{DICES} models. The effect of DICES mainly consists in the broadening of the vibrational peaks and in the lowering of the intensities of the maxima in the spectrum (cf solid and broken curves in Fig.~\ref{fig:spect_8fs1} for all intensities).

\begin{figure}
\includegraphics[scale=0.60]{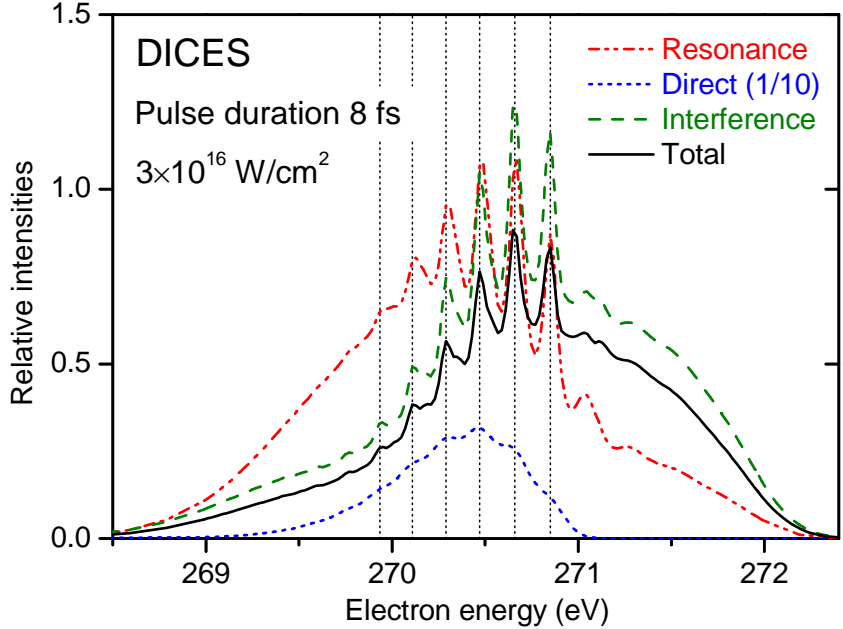}
\caption{(Color online) The RA electron spectra of CO  for an 8 fs Gaussian-shaped pulse of $3\times 10^{16}$~W/cm$^{2}$ peak intensity computed within  the  \emph{DICES}  model (i.e., including rotations). Shown are the  results of  different approximations  discussed in the text. These include the contribution of only the resonant channel  (\emph{Resonant}; dash-dotted curve), of only the direct ionization channel (\emph{Direct}; dotted curve), and of both the direct and resonant channels and the interference between them (\emph{Interference}; dashed curve), as well as the contributions of all mechanisms  including the direct photoionization of the resonance (\emph{Total}; solid curve). Note that the dotted curve is shown on a suppressed scale (factor  $1/10$) compared to the other curves. }
\label{fig:spect_8fs_approx}
\end{figure}

At higher peak intensities (Figs.~\ref{fig:spect_8fs2}--\ref{fig:spect_16fs} for the 8~fs, 4~fs and 16~fs pulses, respectively), a manifold of bifurcations of the line profiles has occurred. In addition, all leakage mechanisms evoked by the strong field contribute significantly to the computed spectra. As a result, the spectra are very different from the experimental one measured in weak fields \cite{Kukk99}. The spectra cover much broader ranges of the electron energies. Very noteworthy is the appearance of the two-hump structure at the low and high energy sides  at the largest field intensities considered here.  It is produced by the strong direct ionization from the ground state, the influence of which grows rapidly with the field intensity \cite{Demekhin11SFatom} (see also the discussion around  Fig.~\ref{fig:spect_8fs_approx} at the end of this subsection).  Obviously, the spectra exhibit many new features. Dramatic difference between the spectra computed within the \emph{no DICES} and \emph{DICES} models can be seen in Figs.~\ref{fig:spect_8fs2}--\ref{fig:spect_16fs}. The spectra computed with and without accounting for the rotational motion (or with and without DICES) differ qualitatively in their structures, possessing different numbers of resonant features at different energy positions, dissimilar widths of peaks and even different relative intensity distributions among them. In some cases (see, e.g., the lower spectra in Figs.~\ref{fig:spect_8fs2} and \ref{fig:spect_16fs}), the positions of local minima and maxima in the spectrum are even alternated.

Let us now discuss the individual contributions of the various leakage mechanisms to the exact RA electron spectra. Fig.~\ref{fig:spect_8fs_approx} depicts the spectra computed within the \emph{DICES} model for the 8~fs pulse at the peak intensity of $3\times 10^{16}$~W/cm$^{2}$. The exact spectrum (\emph{Total}), of course, coincides with that shown in Fig.~\ref{fig:spect_8fs2}. The spectrum computed in the \emph{Resonance} approximation (i.e., in the presence of only RA-decay) is depicted by the dash-dotted curve. As discussed above, at this field intensity several bifurcations of each vibrational peak in the spectrum take place, and the resonant structure of the spectrum is wealthier than that observed in weak fields. The individual contribution of the direct photoionization from the ground state (GS-leakage) is shown in the figure by the dotted curve. This contribution is enormous and the curve is shown on a strongly suppressed scale. Since there is no coupling between the ground and the final ionic states (see Eq.~(\ref{eq:fin2})), the line profile computed in the \emph{Direct} approximation remains unchanged with increasing laser intensity and hence displays no bifurcations. As demonstrated in Ref.~\cite{Demekhin11SFatom}, the intensity of the \emph{Direct} spectrum increases rapidly with the field intensity (see also Fig.~\ref{fig:yield_approx}). If the direct photoionization and resonant spectra were independent of each other, the direct ionization would be the dominant mechanism for the population of the  CO$^+(A~^2\Pi$) final ionic state at this peak intensity (cf dash-dotted and dotted curves in Fig.~\ref{fig:spect_8fs_approx} and note the ${1}/{10}$ factor).

Obviously, since the GS-leakage intensifies with increasing field intensity, the role of the underlying interference effects in the spectrum also grows \cite{Demekhin11SFatom}. At the peak intensity of $3\times 10^{16}$~W/cm$^{2}$, the interference between the resonant and direct channels (\emph{Interference} approximation; dashed curve in Fig.~\ref{fig:spect_8fs_approx}) results in a surprisingly strong suppression of the direct ionization by the competing resonant pathway. The resulting distinct peak patterns in the RA spectrum possess strong asymmetry as seen by comparing the dash-dotted and dashed curves in the figure. With the present parameters utilized in the calculations (both the resonant and the direct electronic amplitudes have the same signs), the interference is more destructive on the low electron energy side than on the high energy side.  At larger intensities, the interference determines not only the shape of the computed RA spectra, but their integral intensities as well (see, e.g., the upper spectra in Figs.~\ref{fig:spect_8fs2} and \ref{fig:spect_4fs}). However, as the total electron yield saturates as a consequence of the GS-leakage mechanism, the impact of the interference effects saturates too \cite{Demekhin11SFatom}.  We note that strong interference effects persist also in the full (\emph{Total}) calculations (solid curve in  Fig.~\ref{fig:spect_8fs_approx}), where the losses of the resonance due to photoionization is additionally taken into account. Depopulation of the resonance due to its ionization into other usually high-lying final ionic states results in an overall suppression of the spectrum without substantially changing its shape (cf dashed and solid curves in the figure). At even higher intensity, however, the peaks seen in the middle of the spectrum will decrease and finally may practically become invisible as a consequence of the strong photoionization of the resonance state.

\subsection{Impact of DICES on vibrations and rotations}
\label{sec:rotations}

\begin{figure}
\includegraphics[scale=0.4]{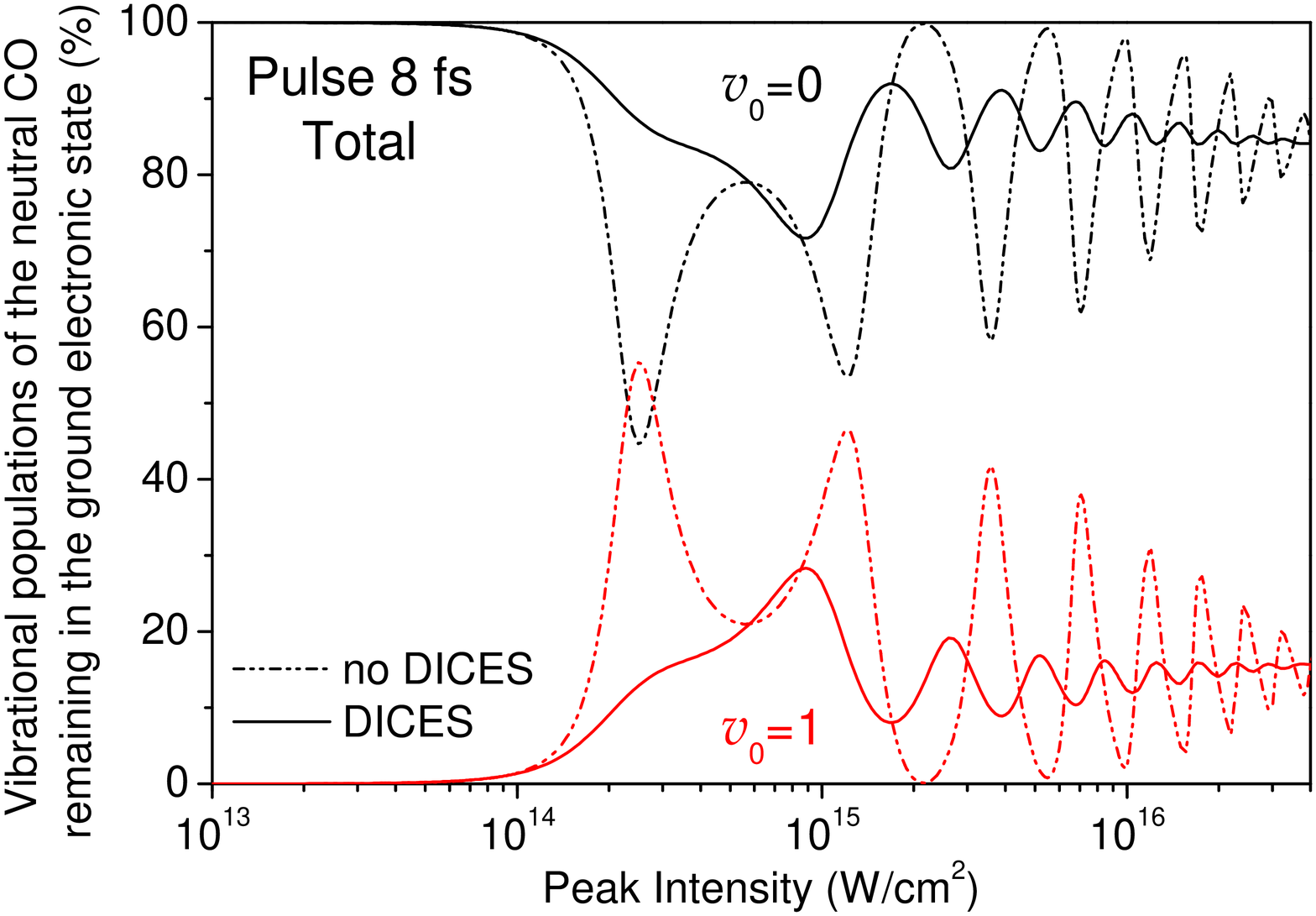}
\caption{(Color online) Impact of the DICES on the vibrational population of the neutral CO molecules remaining in its ground electronic state after the Gaussian-shaped pulse of 8~fs  duration  has expired. Shown are the populations of the $v_0=0$  and 1 vibrational levels as functions of the peak intensity, computed within the \emph{no DICES} (broken curves) and \emph{DICES}  (solid curves) models (\emph{Total}: no approximations). At the considered peak intensities, the population of the $v_0=2$ vibrational level (not shown in the figure) is always well below 1\%. The sum of the populations of all vibrational levels is normalized to 100\% at each peak intensity.}
\label{fig:vibrations}
\end{figure}

\begin{figure}
\includegraphics[scale=0.55]{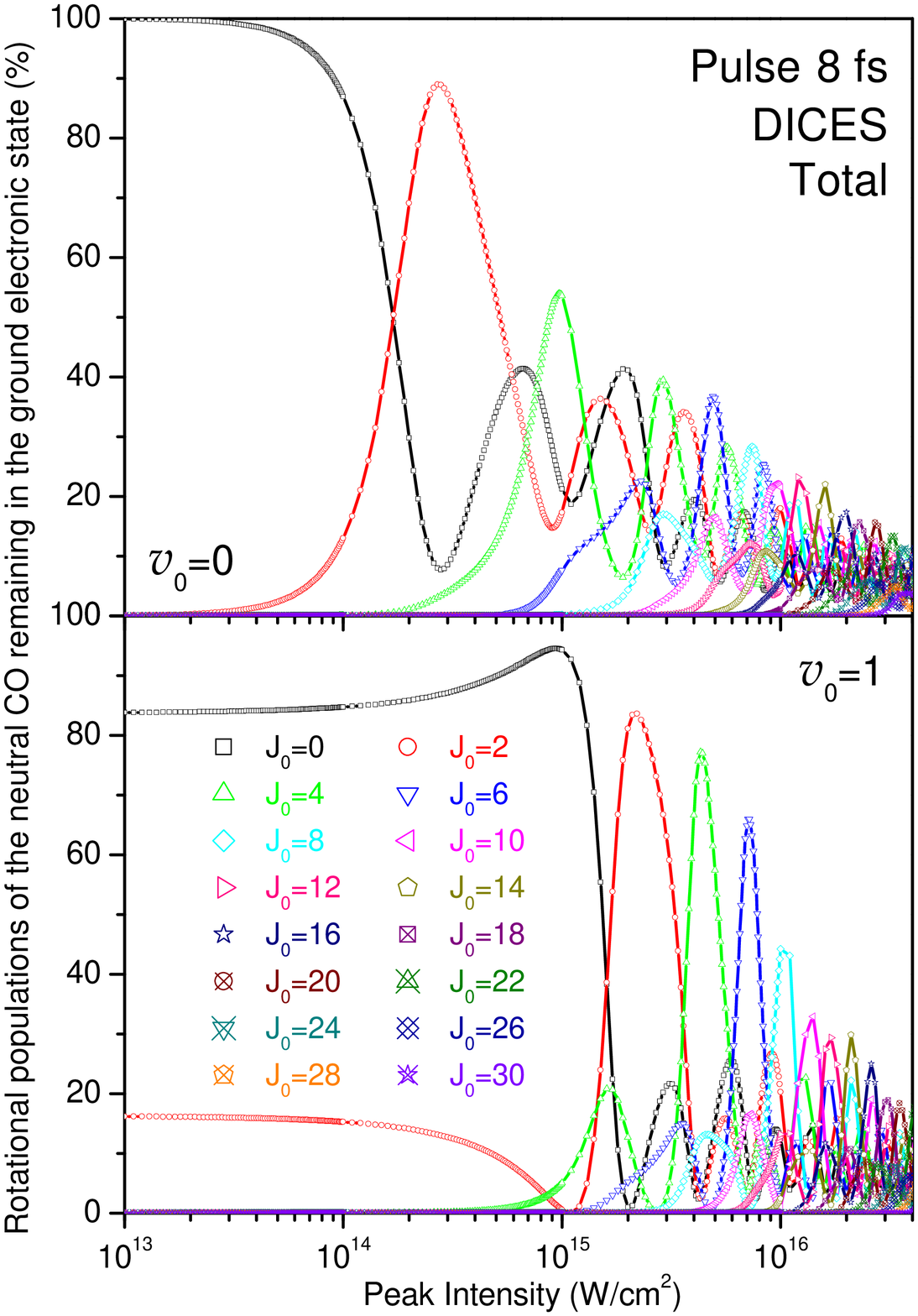}
\caption{(Color online) Impact of the DICES on the rotational population of the neutral CO molecules remaining in its ground electronic state after the Gaussian-shaped pulse  of 8 fs   duration has expired. Shown are the populations of the $J_0$ rotational levels for the $v_0=0$ (upper panel) and $v_0=1$ (lower panel) vibrational levels as a function of the peak intensity computed exactly (i.e., \emph{DICES} model and \emph{Total}). For each vibrational level $v_0$, the sum of the populations of all rotational levels is normalized to 100\% at each peak intensity.}
\label{fig:rotations}
\end{figure}

Before the pulse arrived, the CO molecules have been in their ground ro-vibronic state  CO$(X~^1\Sigma^+,v_0=0,J_0=0)$ which implies that they were isotropically oriented. Fig.~\ref{fig:vibrations} and \ref{fig:rotations} illustrate the change in the vibrational and rotational populations of the neutral molecules remaining in their ground electronic state after the 8~fs  pulse  has expired. In the \emph{no DICES} model, where the rotational motion is excluded, the pulse couples only the electronic and vibrational degrees of freedom and the computed distribution of the neutral CO molecules over the rotational angle $\theta$ remains isotropic represented by only the $J_0=0$ state. However, a part of the energy of absorbed photons is transferred from the electronic excitation to the vibrational ones. As one can see from Fig.~\ref{fig:vibrations} (\emph{no DICES} model; broken curves), at the relatively low intensities below $I_0= 10^{14}$~W/cm$^{2}$ the remaining neutral CO molecules stay mainly in the ground vibrational state $v_0=0$. As the field increases, the excitation of the $v_0=1$ vibrational state is possible. At the peak intensity of around $2.5\times 10^{14}$~W/cm$^{2}$, which correspond to the first maximum in the total electron yield computed in the same model (lower panel of Fig.~\ref{fig:yield_models}), the populations of the $v_0=0$ and 1 vibrational levels become even comparable as can be seen by inspecting the broken curves in Fig.~\ref{fig:vibrations}. At higher intensities, these populations oscillate: the maxima in the population of the $v_0=1$ state  correspond to the maxima in the total electron yield depicted in the lower panel of Fig.~\ref{fig:yield_models} by the dashed curve. We notice that at the considered peak intensities, the population of the $v_0=2$ vibrational level is always below 1\% and, therefore, is not shown in the figure.

Fig.~\ref{fig:rotations} illustrates the changes  in the rotational distributions of the neutral molecules remaining in the $v_0=0$ and 1 vibrational levels of the ground electronic state after the pulse has expired. The impact of the light-induced non-adiabatic effects (DICES), which couples the vibrational and rotational degrees of freedom of the molecule, gives rise to the exchange of a very large number of angular momentum quanta during the ensuing dynamics. As a result, a wide range of   rotational quantum numbers $J_0$ up to about $J_0=30$ is populated in the $v_0=0$ and 1 vibrational levels of the ground electronic state.  The dipole transition selection rule of $\Delta J =\pm 1$ for each single absorbed/emitted photon ensures the population of only even $J_0$ ro-vibronic levels of the neutral molecules remaining in the ground electronic state after the pulse has expired. Fig.~\ref{fig:rotations} shows that higher and higher rotational states $J_0$ are excited with increasing field strength, and the populations of the rotational states possess oscillations with the peak intensity similar to the total electron yield (solid curve in the lower panel of Fig.~\ref{fig:yield_models}). At the relatively low peak intensities below $I_0= 10^{15}$~W/cm$^{2}$, only a few rotational states of small $J_0$ are excited. At higher peak intensities of around $I_0= 10^{16}$~W/cm$^{2}$  the populations of low and high $J_0$ quantum numbers become comparable. The excitation of higher rotational states $J_0$ results in a strongly anisotropic distribution of the neutral molecules over the rotational angle $\theta$ after the pulse is over. We remind that in the \emph{no DICES} model, the distribution over the rotational angle $\theta$ for the ground electronic state remains isotropic as only $J_0=0$ appears. The coupling between the vibrational and rotational degrees of freedom via DICES changes also the vibrational populations of the residual neutral molecules, as demonstrated by the solid curves in Fig.~\ref{fig:vibrations}. The oscillations in the vibrational populations become less pronounced (cf solid and broken curves for each vibrational state $v_0$ in the figure). The maxima in the population of the $v_0=1$ state correspond now to the maxima in the total electron yield computed in the same model (solid curve in the lower panel of Fig.~\ref{fig:yield_models}).

The impact of the DICES on the nuclear dynamics is also documented in the distributions of ions \cite{MolRaSfPRL}, e.g., in the final CO$^+(A~^2\Pi$) electronic state of interest (not shown in the paper). This effect is, however, difficult to measure owing to the following reasons. First of all, these rotational distributions of the final ionic states are different for each energy of the emitted RA electron, and coincidence techniques must be applied. In addition, the ions resulting from the RA decay can be further ionized by the strong pulse, and, as a consequence, taken away from the coincidence signal.

\section{Conclusions}
\label{sec:conc}

The resonant Auger decay of the core-excited C$^\ast$O(1s$^{-1}\pi^\ast,v_r=0$) molecule into the final ionic state CO$^+(A~^2\Pi)$ in an intense X-ray laser field is studied theoretically. The mechanisms responsible for the dynamics of the process are incorporated in the calculations. In particular, we investigated the impact of the light-induced non-adiabatic coupling between the vibrational and rotational degrees of freedom of a diatomic molecule caused by the analogue of the conical intersections of the complex energy surfaces of the ground and `dressed' resonant electronic states (briefly, by the light-induced DICES). The non-adiabatic nuclear wave packet dynamics accompanying the process competes with different ionization mechanisms of the molecule evoked by the laser field: direct ionizations of the ground state (GS-leakage) and of the resonance (RD-leakage), and the Auger decay of the resonance (RA-leakage).

The dynamics of the RA process of a diatomic molecule exposed to strong X-ray pulses is completely different from that predicted for atomic systems, due to the presence of the nuclear  degrees of freedom. In the adiabatic representation, the two-dimensional nuclear wave packet propagates on the complex (owing to the various leakages) potential energy surfaces of the ground and `dressed' resonant states. Being coupled by the non-hermitian interaction induced by the laser field and by the RA decay, these complex surfaces become doubly intersecting in $R$ and $\theta$ space (DICES; doubly intersecting complex energy surfaces). The non-adiabatic couplings at these intersection points are singular, resulting in strong coupling between the electronic, vibrational and rotational degrees of freedom. The imaginary parts of the resulting intersecting complex surfaces, responsible for leakages of the nuclear wave packets from the surfaces, are strongly dependent on both $R$ and $\theta$. As a result, one cannot anymore make a clear separation between the involved leakage mechanisms. Finally, we remind that the finite pulse duration enforces an explicitly time dependent picture of the process.

The individual impacts of the different underlying mechanisms on the dynamics of the process are illustrated on observable quantities like the total electron yield, the RA electron spectra, and the vibrational and rotational distributions of the neutral  molecules remaining in the ground electronic state after the pulse is over. Interference effects and the effects of the competition between the different leakage mechanisms are identified and discussed. These leakage and interference effects increase with the field strength. The non-adiabatic effects result in dramatic changes of the computed total electron yield and electron spectra. These changes are not only of quantitative, but also of unexpectedly strong qualitative nature, and can be verified experimentally. Finally, although the initial rotational distribution has been chosen to be isotropic, the energy exchange between the vibrational and the rotational degrees  of freedom via the light-induced DICES results in strongly non-homogeneous angular distributions of the neutral molecules surviving the pulse and of the ions produced. In the absence of the non-adiabatic effects these distributions would remain isotropic as initially prepared.

The controllability of the light-induced DICES, i.e., of the location of the intersection by laser frequency and of the strength of the interstate coupling by the field intensity, makes the investigation in intense fields of the resonance Auger process and of decay processes in general a challenging and promising new area of research.

\begin{acknowledgments}
The authors would like to thank H.-D. Meyer for his assistance in running the MCTDH package and A.I. Kuleff for many fruitful discussions. Y.--C.C. acknowledges the International Max Planck Research School for Quantum Dynamics in Physics, Chemistry and Biology (IMPRS-QD) for financial support. Financial support by the DFG is gratefully acknowledged.
\end{acknowledgments}


\end{document}